\documentclass[a4paper,10pt,authoryear]{elsarticle}

\usepackage[margin=1in]{geometry}

\usepackage{setspace}
\usepackage{amssymb}
\usepackage{amsmath}
\usepackage{amsthm}

\usepackage{hyperref}
\usepackage{booktabs}
\usepackage{makecell}

\usepackage{color}
\usepackage{endnotes}

\newtheorem{theorem}{Proposition}
\newtheorem*{remark}{Remark}
\renewcommand{\d}{\mathrm{d}}

\journal{}

\begin{document}

\begin{frontmatter}

\title{An analysis of government subsidy policies in vaccine supply chain: Innovation, Production, or Consumption?}

\fntext[fn1]{These authors contributed equally to this work.}
\cortext[cor1]{Corresponding author}

\author[1]{Ran Gu\fnref{fn1}}
\author[1]{Enhui Ding\fnref{fn1}}
\author[2]{Shigui Ma\corref{cor1}}

\ead{sma@nankai.edu.cn}

\affiliation[1]{organization={NITFID, School of Statistics and Data Science, Nankai University},
             city={Tianjin},
             postcode={300071}, 
             country={China}}

\affiliation[2]{organization={College of Tourism and Service Management, Nankai University},
             city={Tianjin},
             postcode={300071}, 
             country={China}}

\begin{abstract}

Vaccines play a crucial role in the prevention and control of infectious diseases. However, the vaccine supply chain faces numerous challenges that hinder its efficiency. To address these challenges and enhance public health outcomes, many governments provide subsidies to support the vaccine supply chain. This study analyzes a government-subsidized, three-tier vaccine supply chain within a continuous-time differential game framework. The model incorporates dynamic system equations that account for both vaccine quality and manufacturer goodwill. The research explores the effectiveness and characteristics of different government subsidy strategies, considering factors such as price sensitivity, and provides actionable managerial insights. Key findings from the analysis and numerical simulations include the following: First, from a long-term perspective, proportional subsidies for technological investments emerge as a more strategic approach, in contrast to the short-term focus of volume-based subsidies. Second, when the public is highly sensitive to vaccine prices and individual vaccination benefits closely align with government objectives, a volume-based subsidy policy becomes preferable. Finally, the integration of blockchain technology positively impacts the vaccine supply chain, particularly by improving vaccine quality and enhancing the profitability of manufacturers in the later stages of production.

\end{abstract}

\begin{keyword}

Supply chain management \sep Game theory \sep Differential games \sep Government subsidy \sep Vaccine supply chain \sep Pricing

\end{keyword}

\end{frontmatter}

\section{Background}

Public health is a critical global issue with far-reaching implications. Improving public health not only enhances individual well-being and safety but also contributes to societal stability and economic development \citep{pantano2020competing}. However, infectious diseases continue to pose a significant challenge to public health efforts. A notable example of this challenge was the 2014 Ebola outbreak in West Africa, which severely strained the healthcare systems of countries such as Guinea, Liberia, and Sierra Leone. The crisis led to over 11,000 deaths and contributed to widespread social unrest and economic hardship\endnote{https://www.who.int/en/news-room/detail/14-01-2016-latest-ebola-outbreak-over-in-liberia-west-africa-is-at-zero-but-new-flare-ups-are-likely-to-occur}. The severity of such public health crises highlights the urgent need for stronger public health awareness and action.

Vaccines represent a major advancement in modern medicine, playing a pivotal role in the prevention and treatment of infectious diseases \citep{vahdani2023fair}. Effective vaccination serves as a powerful defense against these diseases, reducing individual susceptibility and mitigating the risk of transmission \citep{sanae2024towards}. A prominent example of the impact of vaccines is the measles vaccine. Measles, known for its high infection rates and severe consequences, was once a leading cause of child mortality worldwide. Before the widespread introduction of the measles vaccine in 1963, the disease was responsible for an estimated 2.6 million deaths annually\endnote{https://www.who.int/news-room/fact-sheets/detail/measles\label{WHO}}. However, through coordinated vaccination campaigns, many countries have successfully reduced the spread of measles, leading to a significant decline in infection rates. For example, the United States declared the elimination of measles in 2000. According to data from the World Health Organization, between 2000 and 2022, measles vaccination prevented an estimated 57 million deaths\textsuperscript{\ref{WHO}}, positioning it as a cornerstone of public health interventions.

Beyond their medical benefits, vaccines have broader socioeconomic impacts. Vaccination programs reduce the burden on healthcare systems, lower medical costs, promote societal stability, foster economic prosperity, extend life expectancy, and make substantial contributions to global well-being \citep{alam2021challenges}. The importance of vaccines was particularly underscored by the COVID-19 pandemic, which highlighted their critical role and spurred international collaboration in vaccine research to address pressing public health challenges \citep{alam2021challenges}. The pandemic also emphasized the urgent need to strengthen vaccine supply chains, as disruptions in global supply chains revealed its vulnerabilities \citep{ivanov2020viability}.

The successful promotion and distribution of vaccines are essential for ensuring public health \citep{balcik2022mathematical}. Large-scale vaccination campaigns play a vital role in halting disease transmission and achieving herd immunity \citep{wang2023robust}. With an adequate vaccine supply, enhanced distribution networks can ensure accessibility across various regions, resulting in increased coverage and faster vaccination rates \citep{jamison2006disease}. Additionally, government-led health advocacy campaigns can significantly raise public health awareness, boosting vaccination rates and empowering societies to combat diseases \citep{albahri2022novel}. For instance, many governments organize flu vaccination campaigns during the fall season when flu outbreaks are prevalent. These efforts effectively raise community immunity levels and reduce flu transmission. In the fall of 2023, the Beijing government provided free flu vaccinations to vulnerable groups, including schoolchildren and the elderly\endnote{https://www.beijing.gov.cn/zhengce/gfxwj/sj/202307/t20230729\_3210131.html}.

As the use of vaccines becomes increasingly widespread, attention is shifting toward their progress. Recent years have seen remarkable advancements in vaccine manufacturing and technology. Enhanced production capacities and innovative technologies have enabled large-scale vaccine production, ensuring an adequate supply for widespread vaccination campaigns. Simultaneously, the quality of vaccines has steadily improved due to the adoption of advanced technologies. Next-generation vaccines leverage modern biotechnology and manufacturing processes to enhance safety and efficacy while minimizing side effects. For example, improvements in cell culture and purification techniques have strengthened both the production and reliability of the hepatitis B vaccine. The integration of cutting-edge technology platforms, such as mRNA, has accelerated the development process and ushered in a new era of vaccine innovation \citep{anderson2023managing}. Notably, blockchain technology is increasingly being employed in vaccine manufacturing to ensure transparency, credibility, and quality control by combating counterfeit products, enhancing vaccine safety, and enabling traceability \citep{yadav2023blockchain}. In 2020, Zuellig Pharma introduced eZTracker, a blockchain-based tool for tracking medical products, aimed at improving transparency in vaccine distribution and identifying expired or mishandled COVID-19 vaccines\endnote{https://www.zuelligpharma.com/news-insights/Ensuring-safe-vaccines-with-eztracker}. Similarly, in 2021, IBM and Moderna collaborated to create a digital supply chain management platform for COVID-19 vaccines, offering a comprehensive solution for tracking vaccines from production to distribution\endnote{https://newsroom.ibm.com/2021-03-04-Moderna-and-IBM-Plan-to-Collaborate-on-COVID-19-Vaccine-Supply-Chain-and-Distribution-Data-Sharing}. Technological innovation plays a pivotal role in advancing public health, and continuous improvements in vaccine production capacity and quality enhance preparedness for epidemics and safeguard public health.

Vaccines are indispensable to public health, prompting governments worldwide to prioritize and support their supply chains through subsidies \citep{yang2021technology}. These subsidies encompass a range of support mechanisms, including investments in technological advancements aimed at improving vaccine quality and production. For instance, the U.S. Department of Health and Human Services allocated over \$5 billion to Project NextGen, which supports the development of innovative drugs and technologies, including vaccines, to combat COVID-19\endnote{https://aspr.hhs.gov/NextGen}. Recognizing the critical role of vaccine manufacturers in the supply chain, governments frequently implement policies to encourage vaccine development, such as direct subsidies and large-scale vaccine procurement orders \citep{li2022impacts, bian2020tax}. On March 30, 2020, Johnson \& Johnson announced a \$1 billion agreement with the U.S. Department of Health and Human Services to establish a vaccine production line to supply over 1 billion doses. Additionally, based on the delivery of 100 million doses of Ad26.COV2.S, the U.S. government committed to purchasing an additional 200 million doses\endnote{https://www.jnj.com/media-center/press-releases/johnson-johnson-announces-agreement-with-u-s-government-for-100-million-doses-of-investigational-covid-19-vaccine}. Furthermore, governments can subsidize vaccination costs through healthcare insurance to encourage public participation in vaccination campaigns. Specifically, from 2021 to 2022, the Chinese government allocated more than 150 billion CNY from healthcare insurance funds and tax subsidies to provide free or low-cost COVID-19 vaccination services\endnote{http://www.nhsa.gov.cn/art/2023/3/9/art\_7\_10250.html}. Additionally, governments may offer tax incentives to vaccine manufacturers to reduce production costs, stabilize prices, and ensure a steady vaccine supply. These subsidy policies enhance the reliability and sustainability of vaccine supply chains, ultimately strengthening public health protection.

However, limited research has been conducted on the effectiveness and comparison of various government subsidies in vaccine supply chains. This research gap is particularly significant given the potential challenges associated with such policies. Government subsidies may not always achieve the desired outcomes, resulting in inefficiencies within supply chain operations. Despite efforts by governments to promote vaccination through subsidies, issues such as flawed policy design and insufficient collaboration with vaccine manufacturers may hinder successful outcomes \citep{alam2021challenges}. Furthermore, existing studies often overlook the complex dynamics of government subsidy policies within the vaccine supply chain, which could exacerbate the failures of these subsidies. Vaccine manufacturers face ongoing challenges in balancing market demand, pricing strategies, technological investments, and promotional activities \citep{sanae2024towards}. Setting prices too high may reduce public acceptance, while insufficient investment in technology or promotion can undermine vaccine effectiveness and visibility. Conversely, setting prices too low or making excessive investments could result in long-term financial losses for manufacturers. 

Additionally, research gaps persist in understanding manufacturers' perspectives on technology investments, especially through the lens of dynamic game theory. Despite increasing interest in blockchain technology, its application within vaccine supply chains remains underexplored. If these challenges are not addressed promptly, they could disrupt the functioning of vaccine supply chains, complicate vaccine research, and jeopardize the sustainability of manufacturers \citep{alam2021challenges}. Therefore, a comprehensive investigation of the following issues is crucial to promoting the sustainable operation of vaccine supply chains:

\begin{enumerate}
    \item Evaluate the effectiveness and characteristics of various government subsidy methods, identifying the optimal method for specific contexts.
    \item Analyze the optimal decision-making strategies for the government, manufacturers, and retailers under various subsidy schemes, exploring the distinct characteristics of each stakeholder's strategy.
    \item Assess the potential impact of blockchain technology on the vaccine supply chain, with particular emphasis on how changes in key parameters influence decision-making processes.
    \item Investigate opportunities for collaboration between the government and vaccine manufacturers within the supply chain, to identify mutually beneficial outcomes.
\end{enumerate}

To address these questions, we utilized differential equations to model the quality and goodwill of vaccines within the vaccine supply chain. A differential game model was developed, incorporating the key stakeholders: the government, manufacturer, and retailer. This model facilitated the exploration of various government subsidy strategies, including proportional subsidies for technology investments and volume-based subsidies, to determine the optimal approach under different scenarios. Furthermore, under an open-loop condition, we outlined specific optimal decision-making frameworks for all participants in the vaccine supply chain, including mathematical formulations for vaccine pricing and investment strategies. Through detailed analysis and numerical simulations, we examined the characteristics of different subsidy strategies and the impact of key parameters, such as the integration of blockchain technology into the vaccine supply chain. Ultimately, our study provides valuable managerial insights, and the results are expected to fill theoretical gaps in vaccine supply chain research while offering significant guidance for future investigations in related fields.

The paper is organized as follows: Section \ref{sec:lr} provides a review of relevant works on the vaccine supply chain, government subsidy policies, and differential game theory. Section \ref{sec:m} presents our differential game models of the vaccine supply chain under various government subsidy policies, with optimization problems solved under an open-loop assumption. In Section \ref{sec:a}, we primarily discuss the impact of key parameters (e.g., blockchain technology) and variations in government subsidy policies through propositions. Numerical simulations are conducted in Section \ref{sec:ns} to further explore the characteristics of different subsidy policies (e.g., forward-looking versus short-sighted), government subsidy policy decisions, and government-manufacturer cooperation. Section \ref{sec:mi} presents managerial insights derived from these analyses. Conclusions and suggestions for further research are provided in Section \ref{sec:c}. All proofs and supplementary materials are included in the Supplemental online material \ref{app}.

\section{Literature Review}
\label{sec:lr}

\subsection{Vaccine Supply Chain}

The vaccine supply chain has attracted significant attention in recent research, primarily due to its critical role in improving distribution efficiency and driving overall profitability. \citet{duijzer2018literature} conducted a comprehensive review of the existing literature, categorizing research into four main areas: product, production, allocation, and distribution. The findings from these studies are extensive. \citet{wu2005optimization} used stochastic dynamic programming to determine which influenza virus strains should be included in that year's vaccine to match the anticipated pandemic strains. \citet{mohammadi2022bi} developed a robust-stochastic optimization model to minimize death rates and distribution costs in vaccine distribution networks. \citet{pirayesh2025dynamic} proposed a mixed-integer linear programming model to optimize vaccine allocation during pandemics, accounting for age-specific demographic and epidemiological factors.

The existing literature primarily focuses on pricing strategies for established vaccines, with relatively limited research addressing pricing strategies for new vaccines. \citet{robbins2016bilevel} modeled the vaccine pricing problem as a bilevel mathematical programming problem, considering profit-maximizing pricing strategies for a vaccine manufacturer operating in an oligopolistic market. \citet{lee2010pricing} estimated the incremental value and price-demand curves for new vaccines, incorporating factors such as target populations and potential competitors, and derived pricing strategies accordingly. \citet{herlihy2016current} developed an economic model for HPV vaccines, proposing distinct pricing strategies for high- and low-income countries to reduce economic barriers to vaccine uptake while ensuring the profitability of manufacturers. \citet{martonosi2021pricing} addressed the vaccine pricing problem in an oligopolistic market consisting of Pfizer-BioNTech and Moderna, considering optimization and game theory approaches, along with government involvement. \citet{cohen2023price} investigated price discrimination and inventory allocation in the context of vaccines, concluding that regionally discriminatory pricing strategies can be detrimental to firms in competitive markets.

The application of blockchain technology in the vaccine supply chain offers significant potential for enhancing credibility, transparency, and traceability. However, research exploring this emerging technology remains limited. \citet{gao2023vaccine} examined the applicability and effectiveness of blockchain technology through the creation of a two-player dynamic game, showing that blockchain enhances vaccine recipient demand, a finding consistent with our own research. \citet{liu2023strategies} proposed the adoption of blockchain technology under specific conditions, exploring the strategies of vaccine manufacturers and recipients in the integration of blockchain. Adopting a game theory perspective, \citet{liu2021pricing} constructed a vaccine supply chain model to investigate pricing and coordination issues with blockchain adoption and identified the impact of blockchain on the vaccine supply chain.

Similar to \citet{wedlock2019potential}, our research also emphasizes manufacturers' investments in vaccine technology. However, we note that most previous studies focus on isolated aspects, such as vaccine pricing, blockchain technology, or vaccine technology investments. Moreover, existing research on the vaccine supply chain generally adopts a manufacturer-centric perspective. To address these gaps, our study integrates vaccine pricing, blockchain technology, and vaccine technology investments simultaneously through a novel multi-participant, multi-decision-variable model. We identify the optimal decisions for the government, the manufacturer, and the retailer, while assessing how blockchain technology can positively impact and complement existing research.

\subsection{Government Subsidy Policies}

In the current era, the importance of government subsidies in the vaccine supply chain has become increasingly apparent, with the goal of safeguarding public health and ensuring the stability of vaccine distribution. \citet{yang2021technology} explored government financial support for technological advancement through a game model involving two symmetric competing firms, emphasizing the role of subsidies in ensuring sustainability. \citet{xie2023outsourcing} suggested that developed regions should provide appropriate subsidies to local manufacturers to promote re-industrialization and incentivize reshoring for mutual benefit. \citet{bian2020tax} argued that, under certain circumstances, emission reduction subsidies can provide stronger incentives for manufacturers to reduce pollution compared to emission taxes, resulting in increased profitability for all stakeholders.

The majority of existing research has primarily focused on individual forms of government subsidies. \citet{pan2023information} suggested that optimal collaboration between the government and hospitals involves information sharing and subsidies, leading to maximized vaccine coverage and profitability. \citet{xie2021implications} demonstrated that when the research cost coefficient is low or potential demand is high, subsidizing unit production is preferable to subsidizing research and development costs. Conversely, under different conditions, subsidizing research and development costs may be more beneficial. \citet{li2022impacts} examined various government subsidies to manufacturers and retailers in the vaccine supply chain, finding that subsidizing retailers can increase the profits of supply chain members, while subsidies to manufacturers may not have the same effect. \citet{guo2022can} proposed specific subsidy strategies to increase societal vaccine coverage, suggesting that the antivaccine hesitancy effect is only effective when the subsidy is targeted at individuals. Additionally, \citet{pun2021blockchain} advocated for government subsidies to promote the adoption of blockchain technology.

We observe that existing studies predominantly focus on individual forms of government subsidies and lack comprehensive comparisons among multiple types. Our research aims to address this gap by examining various forms of government subsidies, including those for manufacturers' technology investments, volume-based subsidies, and cost reimbursements to vaccine recipients. For each subsidy policy, we derive optimal decisions for each member of the vaccine supply chain and analyze and compare the characteristics of different subsidy policies from the government's perspective.

\subsection{Differential Game Theory}

Differential game theory, a vital branch of game theory, is employed to analyze strategic decision-making and behavior in dynamic settings. Initially introduced by Isaacs \citep{isaacs1999differential} with applications in military strategy, it has since found wide-ranging applications across various fields, including advertising \citep{ji2019monetization}, pricing \citep{liu2016myopic}, food delivery \citep{ma2024cooperate}, and green tourism \citep{ma2021joint}.

In the context of the vaccine supply chain, there are several applications of differential game theory. \citet{gao2023vaccine} explored a vaccine supply chain incorporating artificial intelligence (AI) and blockchain technology. They developed a differential game model featuring inventory levels and concluded that the integration of AI and blockchain technology significantly improves the efficiency and coordination. Similarly, \citet{buratto2020lq} proposed a differential game framework involving both vaccine manufacturers and the healthcare system, identifying the Nash equilibrium strategies for vaccine promotion aimed at increasing vaccination rates.

However, most existing research on the vaccine supply chain tends to approach the problem in a static manner, neglecting the inherently dynamic nature of the game over time. To address this gap, we propose a differential game model for the vaccine supply chain that incorporates dynamic system equations for both vaccine quality and goodwill.

A summary of the relevant literature pertaining to this study is provided in Table \ref{tab: summary}.

\begin{table}[h]
\centering
\caption{Summary of relevant literature}
\setlength\tabcolsep{1.5pt}
\begin{tabular}{ccccccc}
    \toprule
    Authors & \makecell{Vaccine\\supply chain} & Pricing & \makecell{Technology\\investment} & Blockchain & \makecell{Government\\subsidy} & \makecell{Dynamic\\game} \\
    \midrule
    \citet{lin2020cold} & \checkmark & \checkmark &  &  &  & \\
    \citet{martonosi2021pricing} & \checkmark & \checkmark &  &  &  & \\    
    \citet{guo2022can} & \checkmark &  &  &  & \checkmark & \\
    \citet{herlihy2016current} &  & \checkmark & \checkmark &  &  & \\
    \citet{liu2021pricing} & \checkmark & \checkmark &  & \checkmark &  & \\
    \citet{liu2023strategies} & \checkmark & \checkmark &  & \checkmark &  & \\
    \citet{li2022impacts} & \checkmark & \checkmark &  &  & \checkmark & \\
    \citet{pan2023information} & \checkmark & \checkmark &  &  & \checkmark & \\
    \citet{xie2021implications} & \checkmark & \checkmark &  &  & \checkmark & \\
    \citet{xie2023outsourcing} & \checkmark & \checkmark &  &  & \checkmark & \\
    \citet{yang2021technology} & \checkmark & \checkmark & \checkmark &  & \checkmark & \\
    \citet{gao2023vaccine} & \checkmark & \checkmark & \checkmark & \checkmark &  & \checkmark \\
    This paper & \checkmark & \checkmark & \checkmark & \checkmark & \checkmark & \checkmark \\ 
    \bottomrule
\end{tabular}
\label{tab: summary} 
\end{table}
\section{Models}
\label{sec:m}

We analyze a vaccine supply chain involving the government, manufacturer, and retailer. The manufacturer supplies vaccines to the retailer at a retail price, denoted by $\omega$. Additionally, the manufacturer enhances the quality of the vaccine through a technology investment, denoted by $q$, incurring a corresponding cost $C(q)$ for the technology. The retailer, which may be a hospital, vaccination center, or clinic, sells vaccines to customers (i.e., vaccine recipients) at a price $p$, aiming to fulfill the public's vaccination needs. The retailer also enhances the vaccine's goodwill through an advertising investment, denoted by $a$, which incurs a corresponding cost $C(a)$ for advertising. The government provides financial support to participants in the vaccine supply chain through four main funding policies: no subsidy, proportional subsidy for the manufacturer's technology investment, unit subsidy based on vaccine sales volume, and proportional reimbursement for customers' vaccination expenses.

The decisions of each participant in the vaccine supply chain evolve dynamically over time, influencing both the quality and goodwill levels of the vaccines. A summary of the notations used in this study is provided in Table \ref{tab: notation}. 

\begin{table}[h]
\centering
\caption{Notation}  
\begin{tabular}{cc}
    \toprule
    Symbol & Definition\\
    \midrule
    $t$ & Time\\
    $Q$ & Accumulated\ quality\ level\ (State variable)\\
    $G$ & Accumulated\ goodwill\ level\ (State variable)\\
    $p$ & Retail\ price\ (Control variable)\\
    $\omega$ & Wholesale\ price\ (Control variable)\\
    $q$ & Technology\ investment\ of\ manufacturer\ (Control variable)\\
    $b$ & Blockchain\ investment\ of\ manufacturer\ (Control variable)\\
    $a$ & Advertising\ investment\ of\ retailer\ (Control variable)\\
    $D$ & Sales\ volume\\
    $\alpha$ & Market\ capacity\\
    $\beta$ & Price\ sensitivity\ of\ customers\\
    $\theta _1$ & Effects\ of\ technology\ investment\ on\ quality\ level\\ 
    $\theta _2$ & Effects\ of\ blockchain\ investment\ on\ goodwill\ level\\
    $\theta _3$ & Effects\ of\ advertising\ investment\ on\ goodwill\ level\\
    $\gamma _1$ & Effects\ of\ quality\ level\ on\ sales\ volume\\ 
    $\gamma _2$ & Effects\ of\ goodwill\ level\ on\ sales\ volume\\ 
    $\eta$ & Marginal\ sales\ revenue\ of\ government\\ 
    $\delta$ & Decay\ rate\ of\ quality\ and\ goodwill\ level\\
    $r$ & Continuous-time\ discount\ rate\\
    $\pi _G,\pi _M,\pi _R$ & Instantaneous\ profits\ of\ government,\ manufacturer,\ retailer\\
    $V _G,V _M,V _R$ & Accumulated\ profits\ of\ government,\ manufacturer,\ retailer\\
    \bottomrule
\end{tabular}
\label{tab: notation} 
\end{table}

The manufacturer's investments in research and development (R\&D), production, and quality control are crucial for ensuring the quality, safety, and efficacy of vaccines. Consequently, substantial investments in technology typically result in high-quality vaccines. However, a significant reduction or cessation of these investments may lead to a decline in vaccine quality. Such a decline could occur due to factors such as the vaccine lagging behind viral evolution or challenges in maintaining consistent quality control. Therefore, the quality level of the vaccine, denoted as \( Q(t) \), is positively correlated with the manufacturer's technology investment \( q(t) \) and is subject to natural decay, represented by a positive constant \( \delta_1 \). The dynamics of vaccine quality can thus be described by the differential equation in equation \eqref{1}:

\begin{equation}
Q'(t) = {\theta _1}q(t) - \delta_1 Q(t),\quad Q(0) = 0.
\label{1}
\end{equation}

The retailer's investment in advertising plays a crucial role in shaping vaccine goodwill. Advertising enhances the visibility and market appeal of vaccines, significantly contributing to consumer perceptions and confidence in the product. Effective advertising campaigns can increase consumer anticipation and foster a more positive reception of the vaccine. As in the previous context, we model the impact of advertising investment on goodwill using \( \theta_3 a(t) \), as discussed in \citet{ji2019monetization} and \citet{ma2021joint}.

With the rapid advancement of blockchain technology, manufacturers can leverage blockchain to combat counterfeit products, thereby enhancing the transparency and credibility of vaccines while safeguarding their interests \citep{shen2022combating, pun2021blockchain}. Blockchain networks, being transparent, secure, and traceable, offer manufacturers a novel means of demonstrating the vaccine production process and quality control measures, thereby boosting consumer confidence. Additionally, blockchain technology facilitates a deeper understanding of customer feedback, further contributing to the enhancement of the vaccine's reputation. By representing the manufacturer's blockchain investment as \( b(t) \), we draw on the Extension Model presented in \citet{ma2024impacts} and use \( \theta_2 b(t) \) to illustrate the impact of blockchain investment on the goodwill level.

Similarly, a significant reduction or cessation of investment in advertising and blockchain technology can lead to a decline in vaccine goodwill. Contributing factors include natural forgetfulness and the amplification of anti-vaccine sentiments. To account for this, we introduce a positive constant \( \delta_2 \) to denote the natural decay rate of goodwill, allowing us to model the dynamics governing the goodwill level with the differential equation in equation \eqref{2}:

\begin{equation}
G'(t) = {\theta _2}b(t) + {\theta _3}a(t) - \delta_2 G(t),\quad G(0) = 0.
\label{2}
\end{equation}

Finally, we describe the modeling of vaccine sales volume. A decrease in price or an increase in quality or goodwill generally leads to an increase in sales volume. In this study, we attribute the influence on sales volume to both price and non-price factors. The price factor is denoted as \( \alpha - \beta p(t) \), where \( \alpha \) represents market capacity and \( \beta \) represents consumer price sensitivity. This term reflects the proportion of the population willing to accept the current price of the vaccine, given general considerations of its quality and goodwill. The non-price factor, expressed as \( {\gamma _1}Q(t) + {\gamma _2}G(t) \), represents the portion of the population attracted by the vaccine's quality and goodwill. For simplicity, we assume linear effects for all variables, as outlined in \citet{kogan2016learning} and \citet{ma2021joint}. The resulting sales volume can, therefore, be modeled by the equation in \eqref{3}:

\begin{equation}
D(t) = \alpha  - \beta p(t) + {\gamma _1}Q(t) + {\gamma _2}G(t).
\label{3}
\end{equation}

For computational simplicity, we assume that the natural decay rates of both quality and goodwill are identical, denoted by \( \delta > 0 \), as suggested by \citet{zheng2023differential}.

\subsection{No Subsidy Model} 

We begin by considering the scenario in which there is no government intervention. At each time \( t \), the manufacturer sets a wholesale price \( \omega(t) \) for selling vaccines to the retailer, while the retailer sets a retail price \( p(t) \) for selling vaccines to consumers. Accordingly, the sales revenues for the manufacturer and retailer are represented by the first terms of equations \eqref{5} and \eqref{6}, respectively. The government indirectly benefits from vaccination efforts through improvements in public health, reduced disease transmission, and enhanced social stability. Let \( \eta \) denote the government's average marginal revenue per vaccinated individual, with the government's profit described in equation \eqref{4}.

At each time \( t \), the manufacturer determines the technology investment \( q(t) \) and blockchain investment \( b(t) \), while the retailer makes decisions regarding advertising investment \( a(t) \). All investment decisions are constrained to be non-negative. To avoid counterproductive outcomes resulting from excessive investment in any particular area, we impose upper bounds on the control variables to maintain the integrity of the model and ensure its alignment with practical considerations. Specifically, for sufficiently large values of \( q_{\max}, b_{\max}, a_{\max} > 0 \), the following conditions hold: \( 0 \leq q(t) < q_{\max} \), \( 0 \leq b(t) < b_{\max} \), and \( 0 \leq a(t) < a_{\max} \). Consistent with the existing literature \citep{chintagunta1992dynamic, ma2021joint}, we assume that the cost of technology investment follows a quadratic function, given by \( C(q(t)) = \frac{{q^2}(t)}{2} \). This implies that increasing technology investment results in higher costs, i.e., \( \frac{{\partial C(q)}}{{\partial q}} > 0 \), and that the cost of technology investment increases at an accelerating rate, i.e., \( \frac{{\partial^2 C(q)}}{{\partial q^2}} > 0 \). The same principles apply to the costs associated with blockchain and advertising investments.

By incorporating the costs and revenues outlined above, we can derive the instantaneous profit rates for the government, manufacturer, and retailer, as expressed in equations \eqref{4}, \eqref{5}, and \eqref{6}, respectively.

\begin{equation}
    \pi_G = \eta D(t),
    \label{4}
\end{equation}
\begin{equation}
    \pi_M(t) = D(t)\omega(t) - \frac{{q^2}(t)}{2} - \frac{{b^2}(t)}{2},
    \label{5}
\end{equation}
\begin{equation}
    \pi_R(t) = D(t)[p(t) - \omega(t)] - \frac{{a^2}(t)}{2}.
    \label{6}
\end{equation}

In this vaccine supply chain, both the manufacturer and the retailer aim to maximize their accumulated profits over an infinite horizon, as indicated by the integrals in equations \eqref{7} and \eqref{8}, while making decisions simultaneously. Given the continuous-time discount rate \( r \), the optimal control problems for the manufacturer and the retailer can be formulated as follows:

\begin{equation}
    V_M = \max_{\omega > 0, 0 \leq q < q_{\max}, 0 \leq b < b_{\max}} \int_0^\infty \exp(-rt) \pi_M(t) \, \d t,
\label{7}
\end{equation}
\begin{equation}
    V_R = \max_{p > 0, 0 \leq a < a_{\max}} \int_0^\infty \exp(-rt) \pi_R(t) \, \d t.
\label{8}
\end{equation}

Due to the government's involvement, which provides credibility to uphold decisions and enforces compliance with agreed-upon strategies through supervision and other means, we derive an open-loop equilibrium problem. In this problem, strategies are time-dependent functions, which leads us to apply Pontryagin's Maximum Principle for optimization \citep{minner2012dynamic}. The current-value Hamiltonians for the retailer and manufacturer are given in \eqref{9} and \eqref{10}, where $\lambda$ and $\mu$ are the adjoint variables for the retailer and manufacturer, respectively. The Hamiltonian for the retailer's problem in \eqref{9} consists of two parts: the first two terms represent the instantaneous profit derived from \eqref{6}, directly contributing to the objective function in \eqref{8}, while the final term reflects the long-term impact of quality and goodwill levels on profits, contributing indirectly. The adjoint variable $\lambda$ represents the change in $H_R$ when the state variable undergoes a small increase, i.e., the marginal contribution of the state variable. In economic terms, \(\lambda\) is referred to as the shadow price. A similar interpretation applies to \eqref{10}. By maximizing the Hamiltonians, we balance the contributions of both immediate and long-term profits, thereby transforming the dynamic time optimization problem into a static one.

\begin{equation}
    {H_R} = (\alpha - \beta p + \gamma_1 Q + \gamma_2 G)(p - \omega) - \frac{a^2}{2} + \lambda (\gamma_1 Q + \gamma_2 G)',
    \label{9}
\end{equation}
\begin{equation}
    {H_M} = (\alpha - \beta p + \gamma_1 Q + \gamma_2 G)\omega - \frac{q^2}{2} - \frac{b^2}{2} + \mu (\gamma_1 Q + \gamma_2 G)',
    \label{10}
\end{equation}

To ensure the existence and convergence of the solution to the optimization problem, we require that the eigenvalues of the state transition matrix be real numbers and that the stable controls be non-negative. To simplify the problem-solving process, we assume that \( 8\beta \delta (r + \delta) > \Delta \), where \( \Delta = 2 \theta_1^2 \gamma_1^2 + (2\theta_2^2 + \theta_3^2) \gamma_2^2 \).

We use the superscript $*$ to denote the equilibrium solution to the No Subsidy Model. To maintain focus on the key insights and practical implications, detailed expressions for the state and costate variables, along with their corresponding lengthy derivations, are provided in Appendix \ref{app}. For reference, the limits for the state and costate variables are summarized in Table \ref{tab:limit}.

\begin{theorem}
\label{prop:1}
    For the No Subsidy Model:

    (1) The equilibrium prices depend on the quality and goodwill levels, where
    \[
        \omega^* = \frac{{\alpha + \gamma_1 Q^* + \gamma_2 G^*}}{{2\beta}},\quad
        p^* = \frac{{3(\alpha + \gamma_1 Q^* + \gamma_2 G^*)}}{{4\beta}}.
    \]

    (2) The equilibrium efforts depend on the costate variable, which represents the shadow price of the quality and goodwill levels, where
    \[
        q^* = 2\gamma_1\theta_1\lambda,\quad
        b^* = 2\gamma_2\theta_2\lambda,\quad
        a^* = \gamma_2\theta_3\lambda.
    \]

    (3) The state and costate variables take the form \( \Lambda \mathrm{exp}(kx) + B \), where \( \Lambda, k < 0 \) and \( B > 0 \).
\end{theorem}

\subsection{Manufacturer-$q$ Model}

During the pandemic, the government provided subsidies to members of the vaccine supply chain as part of its broader response measures \citep{prentice2020timed}. In the scenario where the government directly provides subsidy support, it subsidizes the manufacturer's technology investment costs at each moment \( t \), in proportion to \( \phi(t) \), paying the manufacturer an amount of \( \frac{\phi(t) q(t)^2}{2} \). Naturally, it is required that \( 0 \leq \phi(t) \leq 1 \). Government-subsidized investments in vaccine technology offer several benefits, including, but not limited to, accelerating vaccine research, improving vaccine quality, and fostering innovation among vaccine manufacturers. The remainder of the framework remains unchanged. The instantaneous profits for the government, manufacturer, and retailer are given by equations \eqref{11}, \eqref{12}, and \eqref{13}, respectively:

\begin{equation}
    \pi_G(t) = \eta D(t) - \frac{\phi(t) q(t)^2}{2},
    \label{11}
\end{equation}
\begin{equation}
    \pi_M(t) = D(t) \omega(t) - \frac{[1 - \phi(t)] q(t)^2}{2} - \frac{b(t)^2}{2},
    \label{12}
\end{equation}
\begin{equation}
    \pi_R(t) = D(t) [p(t) - \omega(t)] - \frac{a(t)^2}{2}.
    \label{13}
\end{equation}

In the three-tier vaccine supply chain, and given the continuous-time discount rate \( r \), the optimal control problems for the government, manufacturer, and retailer can be formulated as equations \eqref{14}, \eqref{15}, and \eqref{16}. We use the superscript \( T \) to denote the equilibrium solution to the Manufacturer-\( q \) Model:

\begin{equation}
    V_G = \max_{0 \leq \phi \leq 1} \int_0^\infty \mathrm{exp}(-rt) \pi_G(t) \, \d t,
    \label{14}
\end{equation}
\begin{equation}
    V_M = \max_{\omega > 0, 0 \leq q < q_{\max}, 0 \leq b < b_{\max}} \int_0^\infty \mathrm{exp}(-rt) \pi_M(t) \, \d t,
    \label{15}
\end{equation}
\begin{equation}
    V_R = \max_{p > 0, 0 \leq a < a_{\max}} \int_0^\infty \mathrm{exp}(-rt) \pi_R(t) \, \d t.
    \label{16}
\end{equation}

\begin{theorem}
\label{prop:2}
    For the Manufacturer-\( q \) Model:

    (1) The equilibrium prices depend on the quality and goodwill levels, where
    \[
        \omega^* = \frac{\alpha + \gamma_1 Q^T + \gamma_2 G^T}{2 \beta}, \quad
        p^* = \frac{3 (\alpha + \gamma_1 Q^T + \gamma_2 G^T)}{4 \beta}.
    \]

(2) The equilibrium efforts and the government's subsidy depend on the costate variables, which represent the shadow prices of the quality and goodwill levels. Specifically, the expressions are given as:

\[
q^{T} = 
\begin{cases} 
\gamma_1\theta_1 \left(\frac{\mu}{2} + \frac{\eta}{4(r + \delta)} \right) & \text{if} \ \phi^T > 0, \\
\gamma_1\theta_1 \mu & \text{if} \ \phi^T = 0,
\end{cases} \quad
b^T = \gamma_2\theta_2 \mu, \quad
a^T = \gamma_2\theta_3 \lambda, \quad
\phi^T = \left[1 - \frac{4(r+\delta) \mu}{\eta + 2(r + \delta) \mu}\right]^+.
\]

(3) When $\eta \ge \frac{4 \alpha \delta (r + \delta)}{8 \beta \delta (r + \delta) - \Delta}$, $\phi^T$ remains positive, and $\mu = 2 \lambda$. In this case, the state and costate variables follow the form $ \Lambda \exp(kx) + B $, where $ \Lambda, k < 0 $ and $ B > 0 $.

\end{theorem}

\subsection{Manufacturer-$D$ Model}

Next, we analyze an alternative form of government subsidy support. In this scenario, at each moment $t$, the government directly subsidizes the manufacturer by an amount $F(t)$ for each individual vaccinated with this brand of vaccine, resulting in a total payment of $D(t)F(t)$ to the manufacturer. It is required that $F(t) \geq 0$. This direct subsidy approach offers several advantages, including the potential to accelerate vaccine production and lower vaccine prices. The instantaneous profits for the government, manufacturer, and retailer are given by the following equations:

\begin{equation}
    \pi_G(t) = [\eta - F(t)] D(t),
    \label{17}
\end{equation}
\begin{equation}
    \pi_M(t) = D(t)[\omega(t) + F(t)] - \frac{q^2(t)}{2} - \frac{b^2(t)}{2},
    \label{18}
\end{equation}
\begin{equation}
    \pi_R(t) = D(t)[p(t) - \omega(t)] - \frac{a^2(t)}{2}.
    \label{19}
\end{equation}

Similar to the Manufacturer-$q$ Model, the optimal control problems for the government, manufacturer, and retailer can be formulated as shown in equations \eqref{20}, \eqref{21}, and \eqref{22}. The superscript $S$ denotes the equilibrium solution for the Manufacturer-$D$ Model.

\begin{equation}
    {V_G} = \max_{F \geq 0}\int_0^\infty  {\exp(- rt)}{\pi _G}(t)\, \d t,
    \label{20}
\end{equation}
\begin{equation}
    {V_M} = \max_{\omega, 0 \leq q < q_{\max}, 0 \leq b < b_{\max}}\int_0^\infty  {\exp(- rt)}{\pi _M}(t)\, \d t,
    \label{21}
\end{equation}
\begin{equation}
    {V_R} = \max_{p, 0 \leq a < a_{\max}}\int_0^\infty  {\exp(- rt)}{\pi _R}(t)\, \d t.
    \label{22}
\end{equation}

\begin{theorem}
\label{prop:3}
    For the Manufacturer-$D$ Model:

    (1) The equilibrium prices and the government's subsidy are functions of the quality and goodwill levels, expressed as:
    \[
        F^S = \left[\frac{{\eta \beta  - (\alpha  + \gamma_1 Q^S + \gamma_2 G^S)}}{{2\beta }}\right]^{+},
    \]
    \[
        \omega^{S} = 
        \begin{cases} 
            \frac{{3(\alpha  + \gamma_1 Q^S + \gamma_2 G^S) - \eta \beta }}{{4\beta }} & \text{if } F^S > 0, \\
            \frac{{\alpha  + \gamma_1 Q^S + \gamma_2 G^S}}{{2\beta }} & \text{if } F^S = 0,
        \end{cases}
        \quad
        p^{S} = 
        \begin{cases} 
            \frac{{7(\alpha  + \gamma_1 Q^S + \gamma_2 G^S) - \eta \beta }}{{8\beta }} & \text{if } F^S > 0, \\
            \frac{{3(\alpha  + \gamma_1 Q^S + \gamma_2 G^S)}}{{4\beta }} & \text{if } F^S = 0.
        \end{cases}
    \]

(2) The equilibrium efforts depend on the costate variables, which represent the shadow prices of the quality and goodwill levels, where  
\[
    q^S = \gamma_1 \theta_1 \mu, \quad b^S = \gamma_2 \theta_2 \mu, \quad a^S = \gamma_2 \theta_3 \lambda.
\]

(3) When $\eta \ge \frac{16\alpha \delta(r + \delta)}{16\beta \delta(r + \delta) - \Delta}$, $F^S$ is always positive, with $\mu = 2\lambda$, and the state and costate variables take the form $ \Lambda \exp(kx) + B $, where $\Lambda$, $k < 0$, and $B > 0$.  
\end{theorem}

To simplify the analysis without sacrificing managerial insights, we assume that the government subsidy is always positive in both models. Details regarding the scenario in which the government may terminate subsidies are provided in \ref{app}.

\begin{table}[h]
    \centering
    \caption{Limits of state and costate variables}
    \begin{tabular}{cccc}
        \toprule
        \textbf{Model} & \textbf{No Subsidy} & \textbf{Manufacturer-$q$} & \textbf{Manufacturer-$D$} \\
        \midrule
        $Q(\infty)$ & $\frac{2\gamma_1 \theta_1^2 \alpha}{8\beta \delta (r + \delta) - \Delta}$ & $\frac{\gamma_1 \theta_1^2}{\delta}\left[\frac{4\alpha \delta (r + \delta) + \eta \gamma_1^2 \theta_1^2}{4(r + \delta)[8\beta \delta (r + \delta) + \gamma_1^2 \theta_1^2 - \Delta]} + \frac{\eta}{4(r + \delta)}\right]$ & $\frac{2\gamma_1 \theta_1^2 (\alpha + \beta \eta)}{32 \beta \delta (r + \delta) - \Delta}$ \\
        $G(\infty)$ & $\frac{\gamma_2 (2\theta_2^2 + \theta_3^2) \alpha}{8\beta \delta (r + \delta) - \Delta}$ & $\frac{\gamma_2 (2\theta_2^2 + \theta_3^2)}{\delta} \cdot \frac{4\alpha \delta (r + \delta) + \eta \gamma_1^2 \theta_1^2}{4(r + \delta)[8\beta \delta (r + \delta) + \gamma_1^2 \theta_1^2 - \Delta]}$ & $\frac{\gamma_2 (2\theta_2^2 + \theta_3^2) (\alpha + \beta \eta)}{32 \beta \delta (r + \delta) - \Delta}$ \\
        $\lambda(\infty)$ & $\frac{\alpha \delta}{8\beta \delta (r + \delta) - \Delta}$ & $\frac{4\alpha \delta (r + \delta) + \eta \gamma_1^2 \theta_1^2}{4(r + \delta)[8\beta \delta (r + \delta) + \gamma_1^2 \theta_1^2 - \Delta]}$ & $\frac{(\alpha + \beta \eta) \delta}{32 \beta \delta (r + \delta) - \Delta}$ \\
        \bottomrule
    \end{tabular}
    \label{tab:limit}
\end{table}

Examining the equilibrium controls, along with the changes in other variables over time, is  crucial for understanding the system's evolution and trends. This is essential for developing more effective strategies. Proposition \ref{prop:4} outlines the time-dependent changes in these variables.

\begin{theorem}
\label{prop:4}
    (1) For the equilibrium solutions to the three models mentioned above, the variables \( q \), \( b \), \( a \), \( Q \), \( G \), \( p \), \( \omega \), and \( D \) all monotonically increase over time.  
    (2) In the Manufacturer-\(q\) model, \( \phi^T \) monotonically decreases over time.  
    (3) In the Manufacturer-\(D\) model, \( F^S \) monotonically decreases over time.
\end{theorem}

Based on Proposition \ref{prop:4}, it is clear that the manufacturer's investment intensity is expected to increase over time. This trend can be attributed to several factors. Initially, the manufacturer may face financial constraints that hinder investment. However, as profits accumulate and the market expands in later stages, the manufacturer is compelled to enhance the quality of the vaccine. Moreover, advanced research is generally more complex than basic technological research, and the costs associated with quality control for high-quality vaccines, represented by the decay rate \( \delta \), are higher. Consequently, greater investment intensity becomes essential for maintaining vaccine quality. A similar logic applies to the retailer. Furthermore, proactive investments by both the manufacturer and the retailer contribute to continuous improvements in vaccine quality and goodwill over time. This progress results in higher vaccine sales volumes, wholesale prices, and retail prices. As a result, a robust vaccine supply chain is well-positioned for sustainable growth. In practice, manufacturers and retailers, as stakeholders, must anticipate these changes and adjust their investment strategies accordingly.

It is also observed that government subsidies to vaccine manufacturers are expected to gradually decrease. In the early stages, the vaccine manufacturer may lack the necessary production technology and experience, requiring substantial subsidies to establish foundational elements such as vaccine technology research and production scale expansion. However, as vaccine manufacturers mature, they become less reliant on government subsidies. Consequently, government subsidies initially serve to protect emerging vaccine manufacturers and promote the overall health and sustainability of the vaccine supply chain. As the industry matures, the government can gradually reduce subsidies, shifting its focus to policies that foster market-driven innovation and competition. This gradual transition ensures that subsidies are used strategically, supporting the growth of emerging manufacturers while encouraging long-term sustainability and efficiency within the vaccine supply chain.

\subsection{Customer-\(p\) Model}

Finally, we analyze a unique form of subsidy support in which the government provides subsidies directly to customers rather than to the manufacturer. In this framework, at each time \( t \), the government reimburses customers for their vaccine expenditures in proportion to \( \psi(t) \), similar to certain healthcare insurance programs. As a result, the total government expenditure is \( \psi(t) p(t) D(t) \). Naturally, it is required that \( 0 \leq \psi(t) \leq 1 \).

Since the actual cost borne by customers for receiving the vaccine is \( [1 - \psi(t)] p(t) \), the price factor in the sales volume model must be amended to \( \alpha - \beta [1 - \psi(t)] p(t) \), as shown in equation \eqref{23}:
\begin{equation}
D(t) = \alpha  - \beta [1 - \psi (t)] p(t) + \gamma_1 Q(t) + \gamma_2 G(t),
\label{23}
\end{equation}
\begin{equation}
{\pi _G} = [\eta -  \psi (t) p(t)] D(t), \quad {V_G} = \max_{0\leq \psi(t) \leq 1} \int_0^\infty  \exp(- rt) {\pi _G}(t)\, \d t,
\end{equation}
\begin{equation}
{\pi _M}(t) = D(t) \omega (t) - \frac{q^2(t)}{2} - \frac{b^2(t)}{2}, \quad {V_M} = \max_{\omega>0, 0\leq q<q_{\max}, 0\leq b<b_{\max}} \int_0^\infty  \exp(- rt) {\pi _M}(t)\, \d t,
\end{equation}
\begin{equation}
{\pi _R}(t) = D(t)[ p(t) - \omega (t)] - \frac{a^2(t)}{2}, \quad {V_R} = \max_{p>0, 0\leq a<a_{\max}} \int_0^\infty  \exp(- rt) {\pi _R}(t)\, \d t.
\end{equation}

For this subsidy strategy to be effective, Proposition \ref{prop:5} suggests otherwise.

\begin{theorem}
\label{prop:5}
    Regardless of how the government determines the subsidy level $\hat\psi(t)$, the manufacturer and retailer always choose
    \[
        \omega (t) = \frac{\omega^* (t)}{{1 - \hat\psi (t)}}, \qquad
        p (t) = \frac{p^* (t)}{{1 - \hat\psi (t)}}.
    \]
\end{theorem}

Proposition \ref{prop:5} demonstrates that, regardless of the level of government subsidies, the sales volume of vaccines remains unchanged compared to the non-subsidized scenario. As a result, the government finds no compelling reason to provide subsidies, leading to the conclusion that $\psi(t) = 0$. This outcome arises because both the manufacturer and the retailer can adjust vaccine prices in response to government subsidies, effectively neutralizing the intended effect on the prices paid by the end consumer. In practice, these funds do not lead to lower consumer prices unless additional safeguards are implemented.

To address this issue and ensure the effective implementation of healthcare coverage, the government must take a more active role in regulating the final price paid by consumers. Potential measures include direct reimbursement to customers based on their actual expenditures, which would help mitigate any price increases. Additionally, the government could prevent price hikes by negotiating pre-agreements with retailers or manufacturers to set clear price limits. Another possible solution would involve imposing penalties on those who attempt to inflate vaccine prices, ensuring that any subsidy provided by the government translates into more affordable vaccines for the public.

These observations highlight the need for robust government intervention in the vaccine supply chain, particularly in balancing the effects of subsidies with the actual price consumers pay. Without such controls, there is a risk that subsidies may be absorbed by manufacturers and retailers rather than benefiting the end consumer, thereby diminishing the intended social impact of these healthcare policies.

\section{Analysis}
\label{sec:a}

This section builds upon our previous findings by conducting a systematic analysis of how changes in system parameters impact equilibrium solutions. Building on this analysis, we compare the magnitudes of steady-state solutions across three different scenarios, synthesizing the results into propositions to enhance our understanding of the roles of parameters and the characteristics of three distinct subsidy strategies. In the following sections, the subsidy policy associated with the Manufacturer-$q$ model will be referred to as Policy (T), due to its emphasis on technology investment, while the subsidy policy corresponding to the Manufacturer-$D$ model will be referred to as Policy (S), due to its focus on sales volume.

\subsection{Impact of Blockchain \& Parameters}

During the COVID-19 pandemic, the proliferation of low-quality and counterfeit vaccines significantly raised public concerns about vaccine safety and reliability. Blockchain technology, with its potential to enhance transparency, traceability, and credibility in the vaccine supply chain, has become a focal point for addressing these issues. By integrating blockchain into vaccine production, manufacturers can bolster public trust and improve the overall vaccine supply chain. However, the adoption of blockchain also presents challenges, including increased investment costs and ongoing maintenance responsibilities. This raises a crucial question: Can blockchain technology generate positive outcomes for vaccine manufacturers within the supply chain? Proposition \ref{prop:6} addresses this question.

\begin{theorem}
\label{prop:6}
    Consider the No Subsidy model. When blockchain technology is introduced (i.e., $\theta _2 > 0$), several changes occur compared to the scenario without blockchain (i.e., $\theta_2 = 0$):\\
    (1) At any given moment, $q$, $b$, $a$, $Q$, $G$, $D$, $\omega$, and $p$ all increase.\\
    (2) After a certain period, $\pi_M$ increases.
\end{theorem}

Proposition \ref{prop:6} indicates that the introduction of blockchain technology by the manufacturer optimally leverages its benefits, such as improving standardization and transparency in production. These advantages contribute to better vaccine quality and an enhanced reputation, which, in turn, lead to increased vaccine sales and stimulate further investments in technology and advertising by both the manufacturer and retailer. However, the implementation of blockchain incurs additional costs, potentially resulting in higher vaccine prices. Initially, the manufacturer may experience a temporary decline in profits due to these additional costs. Nonetheless, as the system becomes more efficient and gains traction, the manufacturer's long-term profitability is expected to improve. For manufacturers with a forward-looking approach, integrating blockchain technology represents a strategically advantageous decision, consistent with the findings of \citet{gao2023vaccine}.

In practice, manufacturers must weigh the initial costs of adopting blockchain technology against the long-term benefits of increased vaccine credibility and sales. Meanwhile, retailers can leverage the increased trust in blockchain-backed vaccines to attract more customers, possibly by adjusting their advertising strategies to highlight the transparency provided by blockchain.

Vaccine supply chains exhibit significant regional variation. In large cities, the vaccine market is typically more substantial due to higher population density, which drives greater demand and sales opportunities. Factors such as infectious disease outbreaks, increased public health awareness, and enhanced vaccine acceptance contribute to market expansion. However, economic conditions marked by low income levels, heightened competition, and a lack of healthcare coverage can result in greater price sensitivity, potentially lowering the acceptable price for vaccines \citep{de2020vaccine}. Moreover, there is growing public attention on vaccine quality and brand reputation, particularly in developed regions or when the safety and efficacy of vaccines are called into question. These factors influence decisions made by both manufacturers and retailers, requiring adaptive strategies in response to changing market dynamics. Accurately estimating these parameters is crucial for optimal decision-making, as discussed in Proposition \ref{prop:7}, which examines how the parameters $\alpha$, $\beta$, and $\gamma_1$ ($\gamma_2$) affect decisions in the absence of subsidies.

\begin{theorem}
\label{prop:7}
    Consider the No Subsidy model. If market capacity $\alpha$ increases, price sensitivity $\beta$ decreases, or the effect of quality (goodwill) level $\gamma_1$ ($\gamma_2$) increases, the following outcomes are observed: \\
    (1) At any given moment, $q, b, a, Q, G, D, \omega, p$ all increase.\\
    (2) After a certain period, $\pi_M$ increases.
\end{theorem}

Proposition \ref{prop:7} indicates that as the focus on vaccine quality and goodwill grows, manufacturers adjust by prioritizing investments in these areas to drive sales growth. High-quality, trusted vaccines can often command higher prices, reflecting their premium status. Additionally, as price sensitivity decreases, manufacturers are more inclined to raise prices to enhance profitability, thereby generating further investment in quality and goodwill efforts. As market capacity expands, manufacturers employ a combination of strategies to maximize long-term profitability. Consequently, the manufacturer's profits steadily increase over time, driven by the successful execution of these investments.

This finding underscores the importance of strategic decisions related to pricing, quality, and goodwill in the vaccine market. Manufacturers must carefully consider the interplay between these factors to optimize their revenue potential. The results suggest that as market conditions improve, manufacturers can increase prices without significantly affecting sales volume, thus leading to higher profitability and greater reinvestment in quality improvement.

The government evaluates the benefit of each individual receiving a vaccine in the supply chain by considering its marginal sales revenue. During epidemic outbreaks, swift action is crucial to control the spread of the virus, break transmission chains, and establish herd immunity. In such critical situations, the government's marginal sales revenue tends to rise. Furthermore, this marginal revenue is notably higher in densely populated areas that are especially vulnerable to outbreaks, as well as in regions with inadequate medical resources to effectively manage the pressures of an epidemic. Therefore, it is vital to treat the government's marginal sales revenue as an important parameter in vaccine policy.

Table \ref{tab:eta} illustrates how marginal sales revenue affects the long-term equilibrium behavior of various variables under different policy scenarios. Analyzing the long-term equilibrium is essential for evaluating the enduring impact of these policies and providing more forward-looking insights for decision-making. The symbols ``\(\nearrow\)'' and ``\(\searrow\)'' denote increases and decreases, respectively.

\begin{table}[h]
    \centering
    \caption{Sensitivity analysis of $\eta$}
    \begin{tabular}{cccccccc}
    \toprule
    Policy & \makecell{$Q(\infty)$, $G(\infty)$, \\ $D(\infty)$} & \makecell{$\pi_G(\infty)$, \\ $\pi_M(\infty)$} & $p(\infty)$ & $\omega(\infty)$ & $\phi^T(\infty)$ & $F^S(\infty)$ \\
    \midrule
    \textbf{Policy (T)} & $\nearrow$ & $\nearrow$ & $\nearrow$ & $\nearrow$ & $\nearrow$ & / \\
    \textbf{Policy (S)} & $\nearrow$ & $\nearrow$ & $\searrow$ \text{, when } $4\beta\delta(r+\delta) > \Delta$ & $\searrow$ & / & $\nearrow$ \\
    \bottomrule
    \end{tabular}
    \label{tab:eta}
\end{table}

The results presented in Table \ref{tab:eta} are consistent with empirical observations. As marginal sales revenue increases, the government is incentivized to intensify its efforts to promote vaccination, which, in turn, expands the market demand for vaccines. Concurrently, with increased financial resources, the government typically enhances subsidies to manufacturers, encouraging them to expand production. This strategy not only ensures an adequate vaccine supply but also provides manufacturers with stable distribution channels and reliable revenue streams, further incentivizing them to improve vaccine quality. These findings align with those of \citet{yang2021technology} and \citet{li2022impacts}.

Moreover, when the government subsidizes manufacturers for vaccine technology investments, the increased quality and production costs associated with high-quality vaccines lead to higher prices. In contrast, when the government offers subsidies based on sales volume, this typically results in lower vaccine prices, as manufacturers seek to increase sales and capitalize on the subsidies, especially when the public is price-sensitive.

\subsection{Comparison among Different Subsidy Policies}

The implementation of various subsidy policies yields different effects on the vaccine supply chain. In analyzing the characteristics and differences among these policies, it is essential to focus on the government's stable profits. Stable profits reflect the overall impact of the subsidy policy on both financial outcomes and public health, providing insight into its effectiveness from the government's perspective. Typically, the government opts for a subsidy policy that maximizes its revenue. To assist in this decision-making process, we conduct a comparative evaluation of the government's stable profits under different subsidy policies, offering valuable insights for policy selection. This analysis is presented in Proposition \ref{prop:9}.

\begin{theorem}
\label{prop:9}
    In the three scenarios, the stable profits of the government follow the relationships:\\
    (1) ${\pi _G^{T}}(\infty) \geq {\pi _G^{*}}(\infty)$.\\
    (2) When $\eta$ is sufficiently large, ${\pi _G^{S}}(\infty)>{\pi _G^{*}}(\infty)$.\\
    (3) When $\eta$ is sufficiently large, suppose $r \rightarrow 0^+$ and $\gamma_1\theta_1=\gamma_2\theta_2=\gamma_2\theta_3=\rho$. When $\beta > 1.04462\dfrac{\rho^2}{\delta^2}$, ${\pi _G^{S}}(\infty)>{\pi _G^{T}}(\infty)$. Otherwise, ${\pi _G^{T}}(\infty) \geq {\pi _G^{S}}(\infty)$.
\end{theorem}

From Proposition \ref{prop:9}, it follows that if the government uses stable profits as the criterion for policy design, Policy (T) is preferable to the absence of a subsidy policy. However, the choice between Policy (S) and Policy (T) depends on specific conditions. When marginal sales revenue \(\eta\) is sufficiently large, Policy (S) may outperform Policy (T); however, this outcome is contingent on factors such as market sensitivity and the government's discount rate. This nuance in policy evaluation underscores the importance of making tailored decisions based on real-time economic conditions and epidemic circumstances. A more detailed discussion of this aspect will be presented in Section \ref{sec:ns}.

The stable profit of the manufacturer is a key indicator in evaluating the effects of subsidy policies. It reflects the manufacturer's market performance, with higher profits supporting vaccine research and market expansion. Comparing stable profits allows for an assessment of the level of support provided by different policies, highlighting those that best promote the manufacturer's sustainable development. Proposition \ref{prop:10} summarizes the comparison of the manufacturer's stable profits under various subsidy policies.

\begin{theorem}
\label{prop:10}
    In the three scenarios, the stable profits of the manufacturer follow the relationships:\\
    (1) ${\pi _M^{T}}(\infty) \geq {\pi _M^{*}}(\infty)$.\\
    (2) When $\eta$ is sufficiently large, ${\pi _M^{S}}(\infty)>{\pi _M^{*}}(\infty)$.\\
    (3) When $\eta$ is sufficiently large, suppose $r \rightarrow 0^+$ and $\gamma_1\theta_1=\gamma_2\theta_2=\gamma_2\theta_3=\rho$. When $\beta > 0.68255\dfrac{\rho^2}{\delta^2}$, ${\pi _M^{S}}(\infty)>{\pi _M^{T}}(\infty)$. Otherwise, ${\pi _M^{T}}(\infty) \geq {\pi _M^{S}}(\infty)$.
\end{theorem}

According to Proposition \ref{prop:10}, when considering only the manufacturer's stable profit, Policy (T) is preferable to a no-subsidy policy. However, the preference for Policy (S) is contingent upon specific conditions.

It is evident that Proposition \ref{prop:10} closely aligns with Proposition \ref{prop:9}, indicating that the optimal subsidy policy for the government often mirrors the manufacturer's strategy. This convergence arises from the shared objectives of both parties: the government benefits from increased vaccine sales and enhanced public immunization coverage, while the manufacturer maximizes profits through expanded sales. The alignment between the two stakeholders fosters a mutually beneficial environment, which not only supports short-term vaccine distribution but also promotes long-term vaccine research and the sustainability of the entire supply chain. This finding underscores the potential for government subsidy policies to be structured in ways that not only advance public health but also incentivize vaccine manufacturers to invest in quality improvements and research. Policymakers may, therefore, consider aligning subsidy levels with pricing structures that benefit manufacturers, ensuring that both parties work toward the common goal of public immunization while sustaining the economic health of the vaccine supply chain.

In our analysis, we introduce the variable $A(t) = \gamma_1 Q(t) + \gamma_2 G(t)$ as the aggregate level of quality and goodwill. This formulation allows us to focus on the combined impact of non-price factors on sales volume. By doing so, we capture the simultaneous effects of changes in both quality and goodwill on demand in a balanced manner.

To assess how government subsidy policies influence the investments of the manufacturer and retailer, as well as the quality and goodwill levels of vaccines, we compare the steady-state values of these investments and aggregate levels. This comparative analysis is summarized in Propositions \ref{prop:11} and \ref{prop:12}.

\begin{theorem}
\label{prop:11}
    In the three scenarios, the stable controls of the investments follow the relationships: \\
    (1) $q^{T}(\infty) \geq q^{*}(\infty),\ b^{T}(\infty) \geq b^{*}(\infty),\ a^{T}(\infty) \geq a^{*}(\infty)$. \\
    (2) $q^{T}(\infty) > q^{S}(\infty)$.
\end{theorem}

\begin{theorem}
\label{prop:12}
    In the three scenarios, the stable state of the aggregate level follows the relationships: \\
    (1) $A^{T}(\infty) \geq A^{*}(\infty)$. \\
    (2) Let $n = \dfrac{\Delta}{8 \beta \delta (r + \delta)} \in (0, 1)$, if $\dfrac{6 - n}{1 - n} > \dfrac{\gamma_2^2 (2\theta_2^2 + \theta_3^2)}{\gamma_1^2 \theta_1^2}$, then $A^{T}(\infty) > A^{S}(\infty)$.
\end{theorem}

\begin{remark}
    The condition stated in Proposition \ref{prop:12}(2) is generally satisfied in practical applications, as the effect of technological effort tends to be relatively significant. It can be relaxed to $6 \gamma_1^2 \theta_1^2 > \gamma_2^2 (2\theta_2^2 + \theta_3^2)$.
\end{remark}

Propositions \ref{prop:11} and \ref{prop:12} suggest that, compared to the non-subsidized scenario, both the manufacturer and retailer are likely to increase their investment efforts under Policy (T). Specifically, direct government subsidies to the manufacturer for technology reduce cost pressures, thereby significantly enhancing the manufacturer's technological investments and improving the quality and goodwill of vaccines. While Policy (S) does not guarantee increased investments in blockchain or advertising, the strong emphasis on technology under Policy (T) ensures an increase in the overall level of vaccine quality and goodwill.

To evaluate the impact of government subsidy policies on the pricing strategies of the manufacturer and retailer, as well as on the sales volume of vaccines, it is necessary to compare both the initial and steady-state values of prices and sales. This comparative analysis is summarized in Propositions \ref{prop:13} and \ref{prop:14}.

\begin{theorem}
\label{prop:13}
    In the three scenarios, the initial and stable price values follow the relationships:

        (1) $p^{S}(\tau) < p^{*}(\tau) < p^{T}(\tau)$, $\omega ^{S}(\tau) < \omega ^{*}(\tau) < \omega ^{T}(\tau)$, where $\tau > 0$ is sufficiently small.

        (2) $p^{S}(\infty) < p^{*}(\infty) \leq p^{T}(\infty)$, $\omega ^{S}(\infty) < \omega ^{*}(\infty) \leq \omega ^{T}(\infty)$.
\end{theorem}

\begin{theorem}
\label{prop:14}
    In the three scenarios, the initial and stable sales volume values follow the relationships:

        (1) $D^{S}(\tau) > D^{T}(\tau) > D^{*}(\tau)$, where $\tau > 0$ is sufficiently small.

        (2) $D^{T}(\infty) \geq D^{*}(\infty)$.

        (3) When $\eta > \dfrac{{\alpha \left( 16\beta \delta \left( r + \delta \right) + \Delta \right)}}{{2\beta \left( 8\beta \delta \left( r + \delta \right) - \Delta \right)}}$, $D^{S}(\infty) > D^{*}(\infty)$.
\end{theorem}

Propositions \ref{prop:13} and \ref{prop:14} indicate that under Policy (S), where the government provides direct subsidies based on sales volume, both the manufacturer and retailer can increase sales by lowering vaccine prices. This strategy is particularly effective in the early stages, before significant vaccine quality and goodwill are established. In contrast, under Policy (T), which involves higher investment costs for vaccine development, both the manufacturer and retailer respond by increasing vaccine prices, leading to lower initial sales volumes. However, as vaccine quality and goodwill improve over time, sales under Policy (T) are expected to increase rapidly, and may potentially surpass those under Policy (S).

\section{Numerical Simulation}
\label{sec:ns}

Given the complexity of the model, it is challenging to present the nature and comparison of certain variables solely through propositions. Therefore, we conduct numerical simulations to provide a more intuitive analysis of the research findings, focusing on the characteristics of the two subsidy policies and the cooperation between the government and the manufacturer. Based on relevant literature \citep{ma2021sustainable, gao2023vaccine} and reasonable assumptions, we set the following parameter values: $\alpha = 18, \, \beta = 7, \, \theta_1 = 1, \, \theta_2 = 0.5, \, \theta_3 = 0.8, \, \gamma_1 = 0.3, \, \gamma_2 = 0.2, \, \eta = 7, \, \delta = 0.1, \, \text{and }r = 0.03$.

\subsection{Characteristics of Subsidy Policies}

In this subsection, we perform numerical simulations to compare the magnitudes of various variables and examine the characteristics of different subsidy policies.

\begin{figure}[!h]
    \centering
    \includegraphics[scale = 0.27]{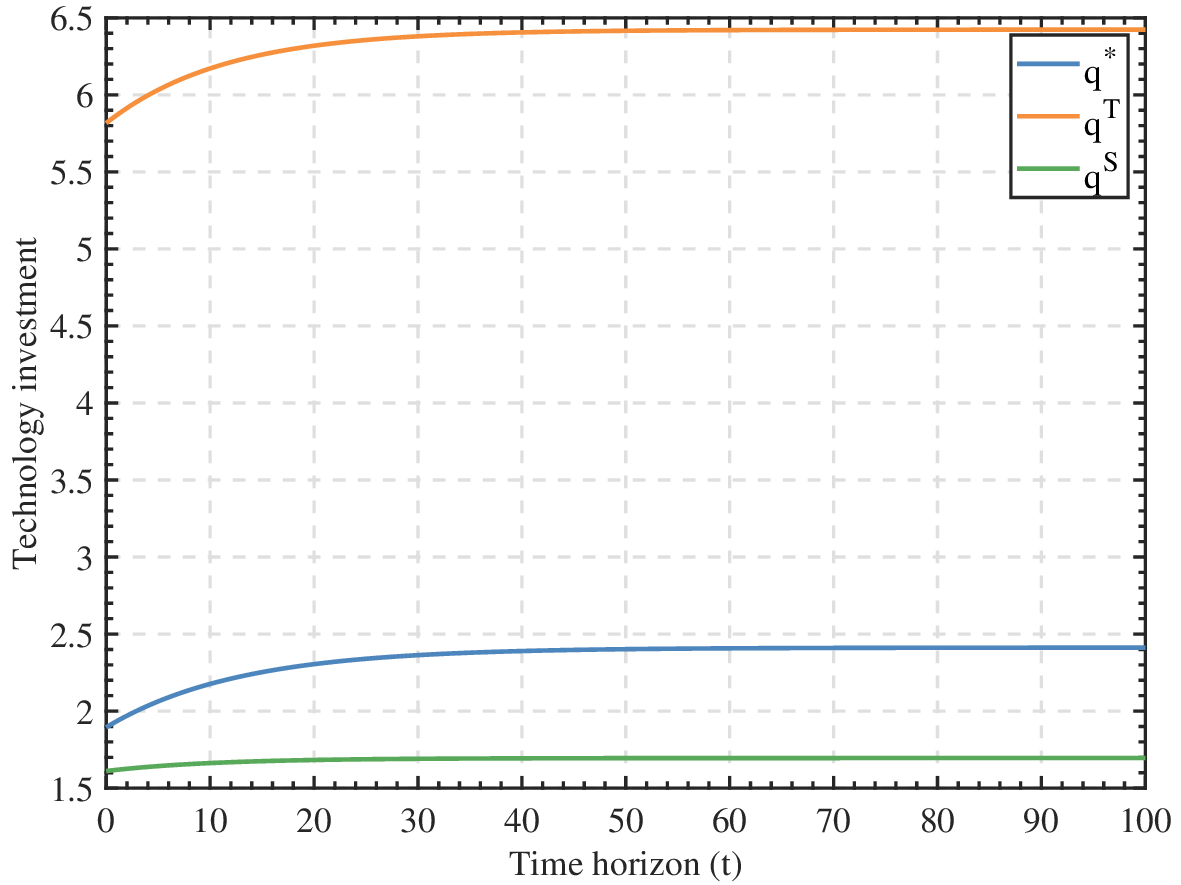}
    \qquad
    \includegraphics[scale = 0.27]{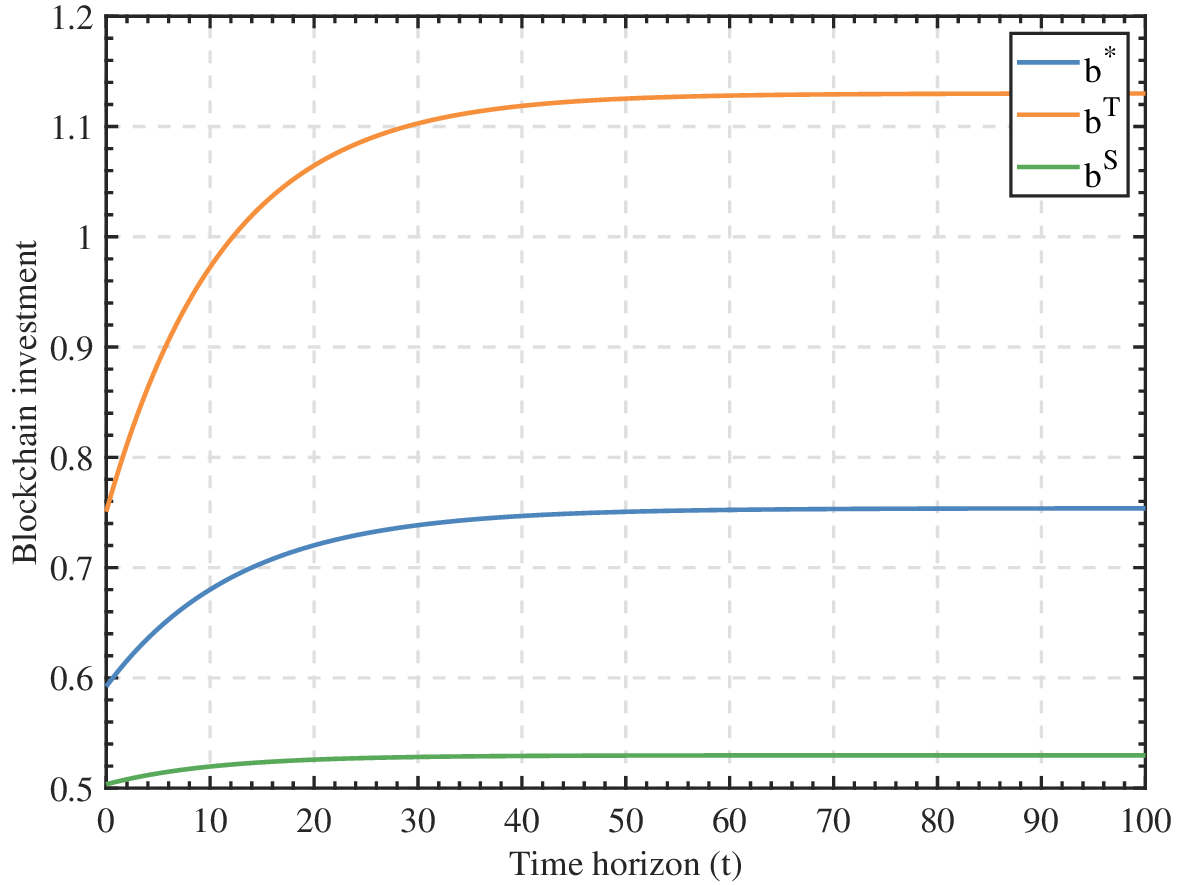}
    \par\bigskip
    \includegraphics[scale = 0.27]{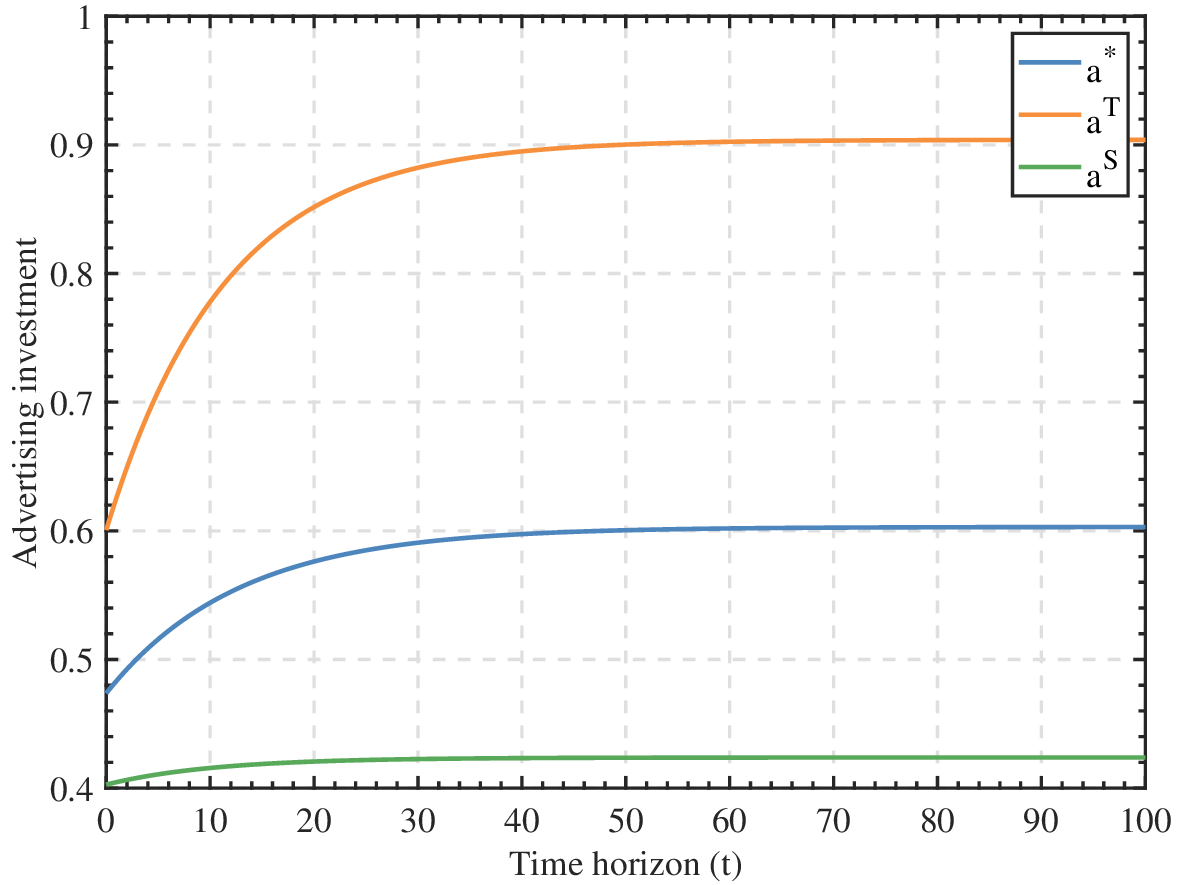}
    \qquad
    \includegraphics[scale = 0.27]{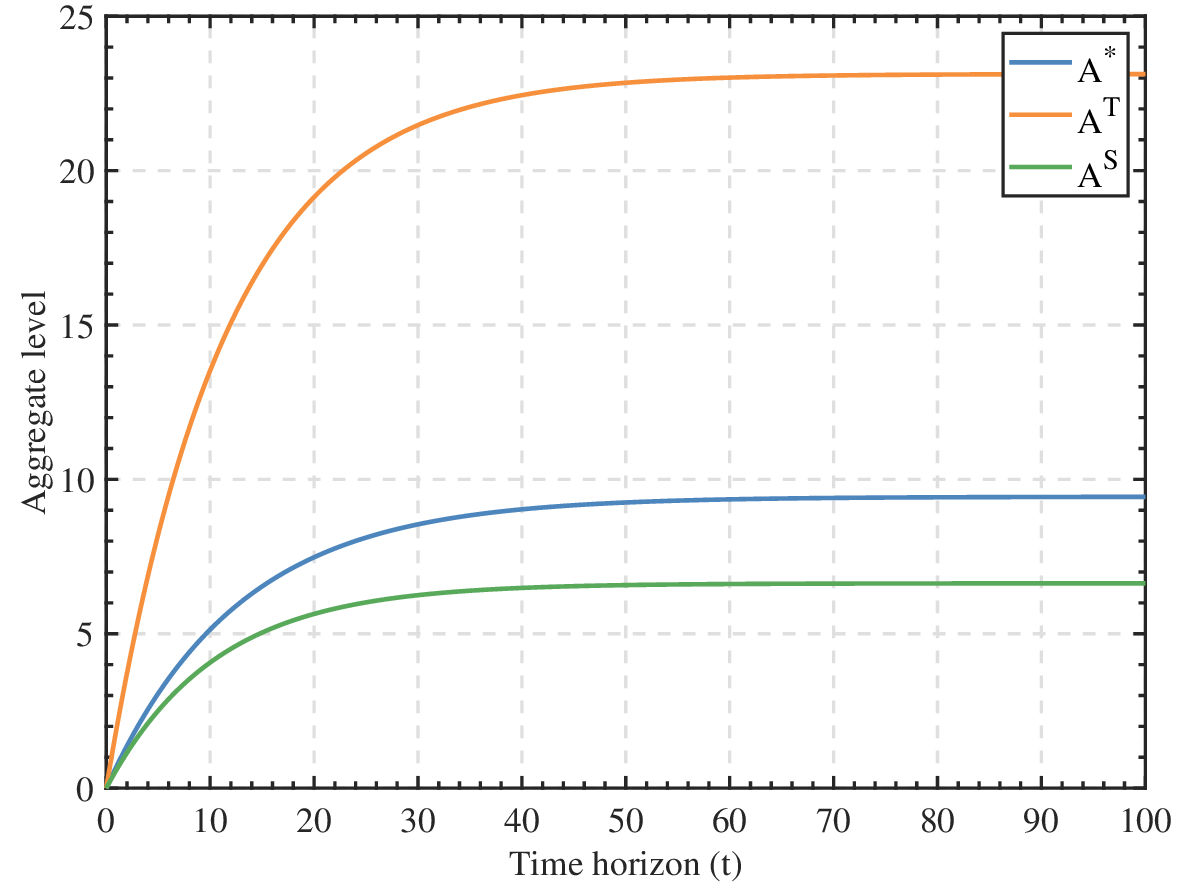}
    \caption{Comparisons of $q$, $b$, $a$, and $A$ under various scenarios.}
    \label{fig:1}
\end{figure}

\begin{figure}[!h]
    \centering
    \includegraphics[scale = 0.27]{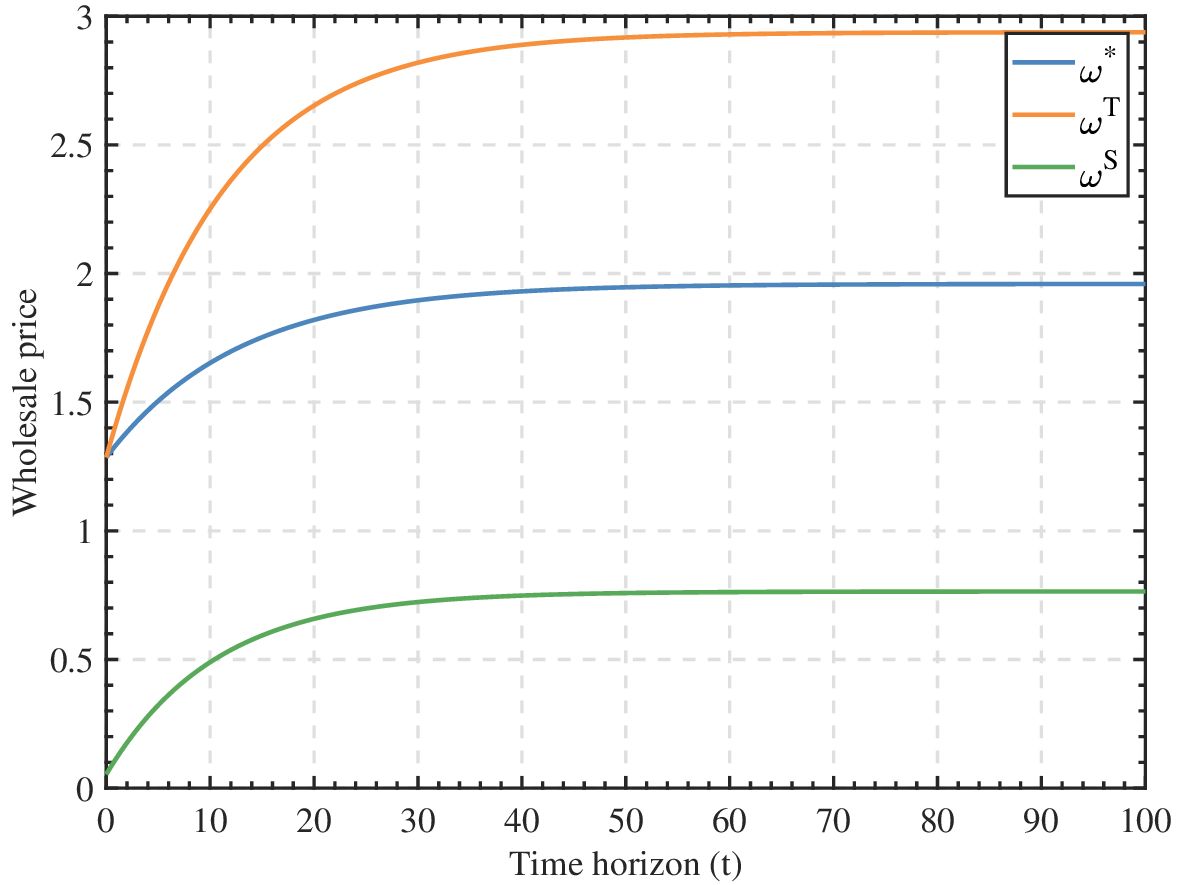}
    \qquad
    \includegraphics[scale = 0.27]{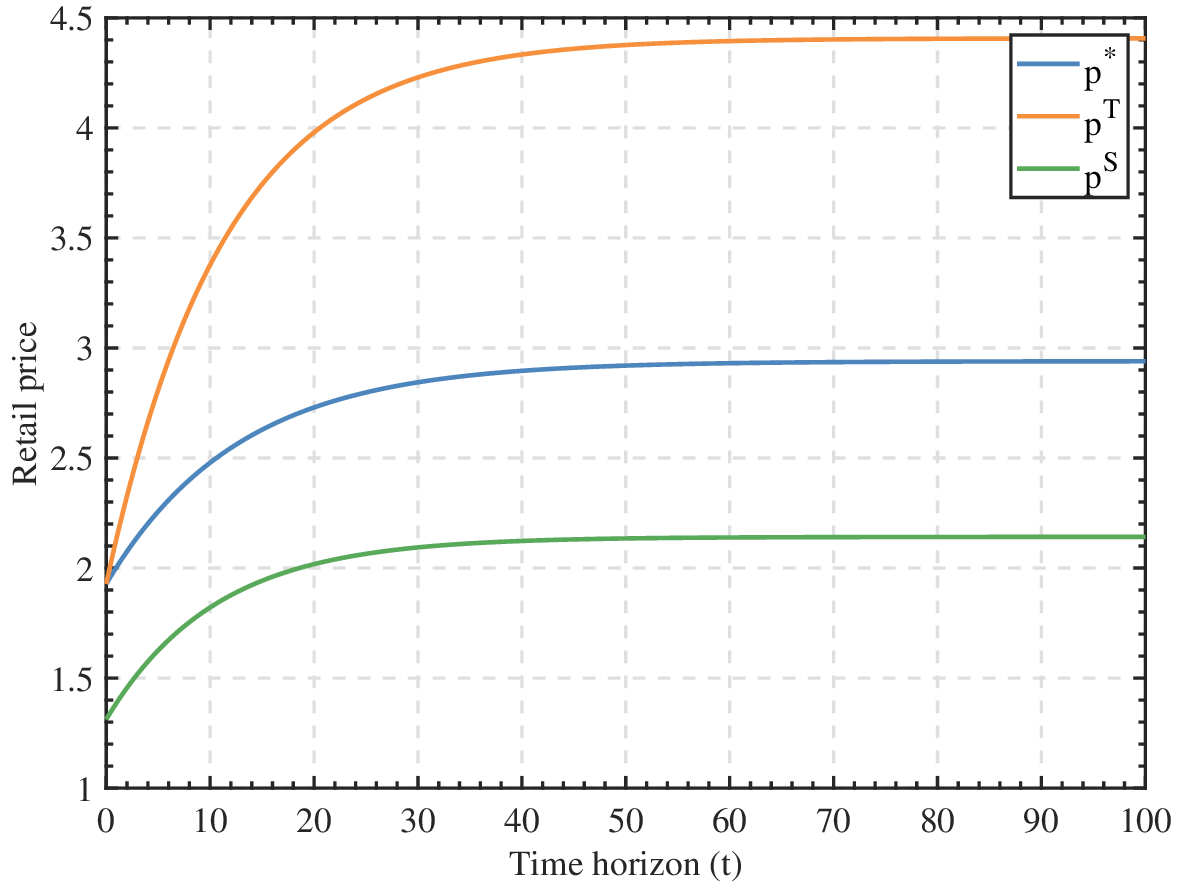}
    \caption{Comparisons of $\omega$ and $p$ under various scenarios.}
    \label{fig:2}
\end{figure}

\begin{figure}[!h]
    \centering
    \includegraphics[scale = 0.27]{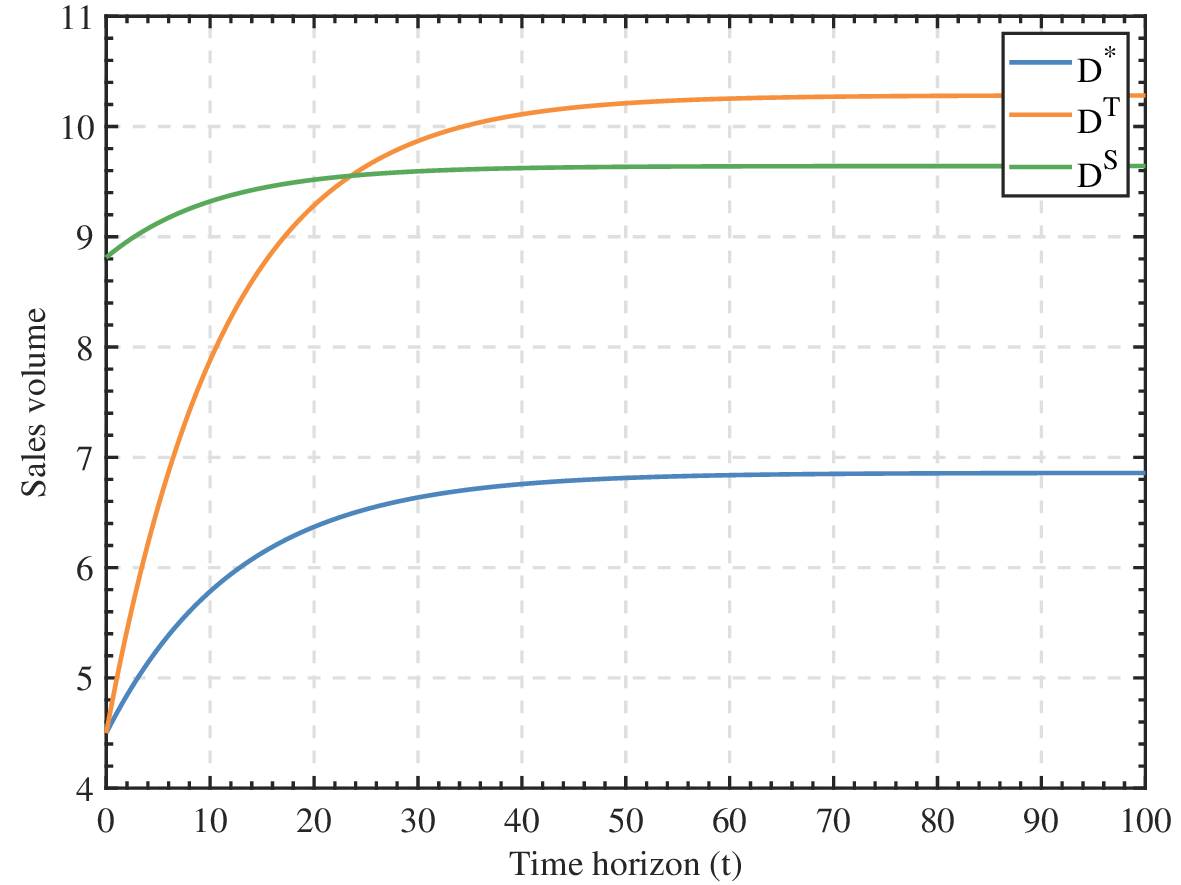}
    \qquad
    \includegraphics[scale = 0.27]{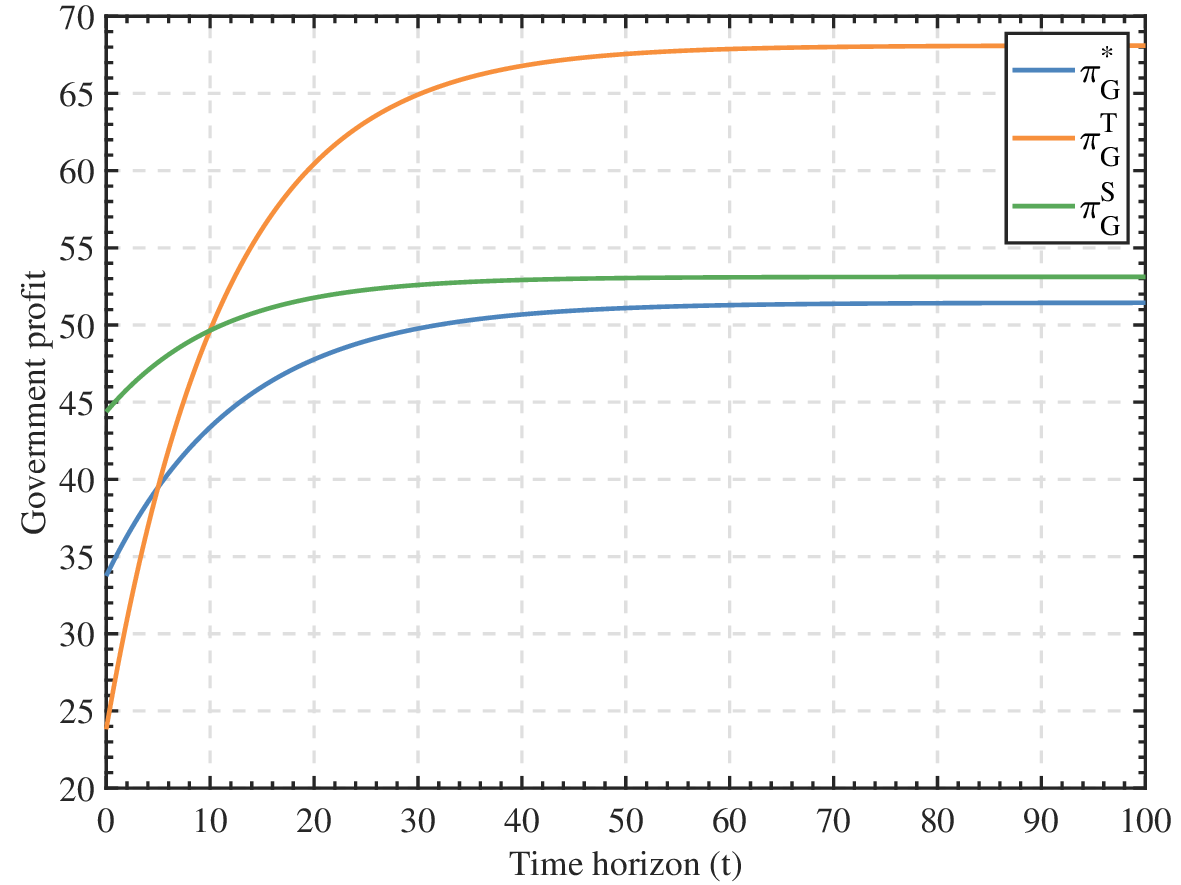}
    \caption{Comparisons of $D$ and $\pi_G$ under various scenarios.}
    \label{fig:3}
\end{figure}

As shown in Figure \ref{fig:1}, under Policy (T), both the manufacturer's and retailer's investments consistently remain higher, particularly in terms of the manufacturer's technology investment. In contrast, under Policy (S), the opposite occurs, resulting in a significantly higher aggregate investment under Policy (T) compared to Policy (S). Figure \ref{fig:2} illustrates that vaccine prices are consistently higher under Policy (T) compared to the lower prices observed under Policy (S). Furthermore, Figure \ref{fig:3} demonstrates that in the early stages, the government achieves higher sales and profits under Policy (S) than under Policy (T). However, in the later stages, Policy (T) leads to higher sales and profits. This comparative analysis is summarized in Table \ref{tab:compare}, where $\tau > 0$ is considered sufficiently small. In Table \ref{tab:compare}, a ``$+$'' indicates a higher value compared to the other policy, while a ``$-$'' indicates a lower value.

\begin{table}[h]
    \centering
    \caption{Characteristics of Two Subsidy Policies}
    \begin{tabular}{cccccccc}
        \toprule
        Policy & $q$, $b$, $a$ & $A$ & $\omega$, $p$ & $D(\tau)$ & $D(\infty)$ & $\pi_G(\tau)$ & $\pi_G(\infty)$ \\
        \midrule
        \textbf{Policy (T)} & $+$ & $+$ & $+$ & $-$ & $+$ & $-$ & $+$ \\
        \textbf{Policy (S)} & $-$ & $-$ & $-$ & $+$ & $-$ & $+$ & $-$ \\
        \bottomrule
    \end{tabular}
    \label{tab:compare}
\end{table}

The observed trends indicate that Policy (T) emphasizes the development of high-quality, reputation-driven vaccines through sustained and significant investments by both manufacturers and retailers. This strategy aims to enhance the overall quality and goodwill of the vaccines. In the initial phase, higher prices lead to lower sales volumes, while substantial investment costs reduce profits. However, in the later stages, the increased sales volume of vaccines results in higher future profits for the government, making Policy (T) a forward-looking approach. In contrast, Policy (S) focuses on boosting vaccine sales through continuous price reductions initiated by both manufacturers and retailers. This approach enables the government to generate higher short-term profits. However, due to the limited accumulation of vaccine quality and goodwill, the sales volume is lower compared to Policy (T). Consequently, the government's profits in the later stages are also reduced, making Policy (S) a relatively short-term strategy. Since early profits sometimes carry greater weight than future profits, the government must carefully evaluate the prevailing circumstances to choose the most appropriate subsidy policy.

Based on the insights derived from these subsidy policies, a key recommendation is that the choice of policy should align with the strategic goals. If the objective is to rapidly increase vaccine accessibility and market penetration, Policy (S) may be more suitable due to its focus on price reductions and immediate sales growth. However, if the goal is to establish a high-quality product with long-term consumer trust, Policy (T) would be the more effective choice, as it supports sustained investment in vaccine quality and goodwill.

\subsection{Subsidy Policy Selection and Member Collaboration}

\begin{figure}[!h]
    \centering
    \begin{minipage}{0.65\linewidth}
    \centering
    \includegraphics[scale = 0.3]{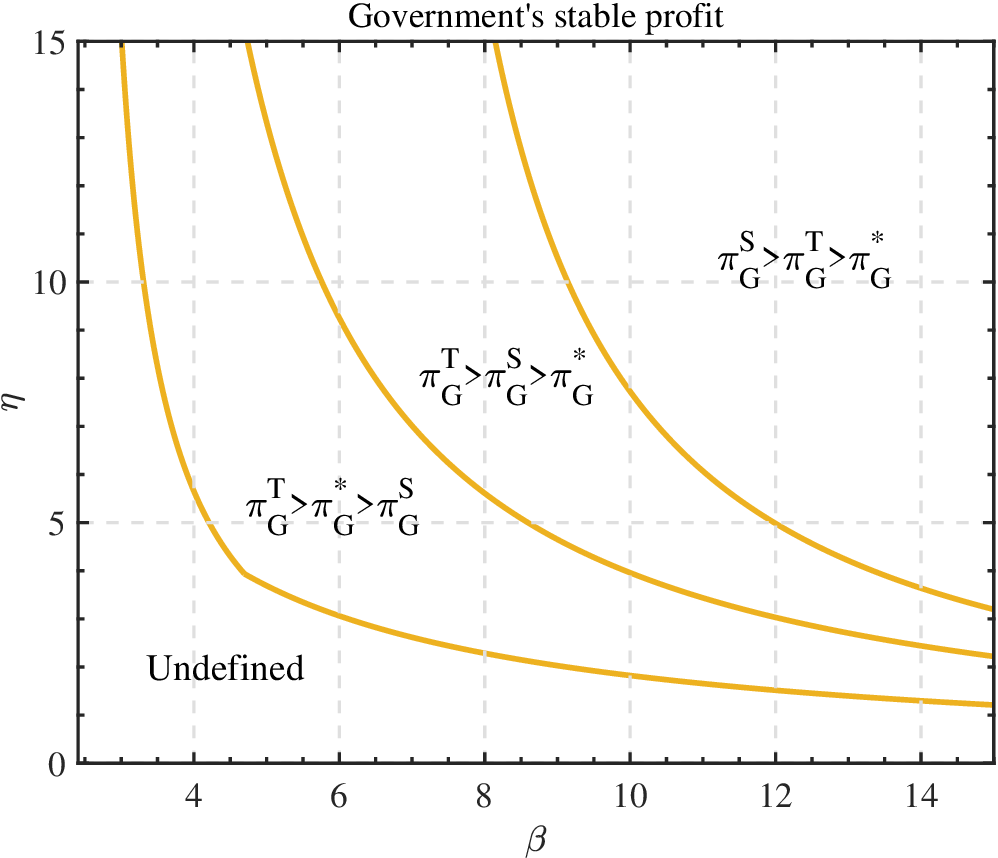}
    \includegraphics[scale = 0.3]{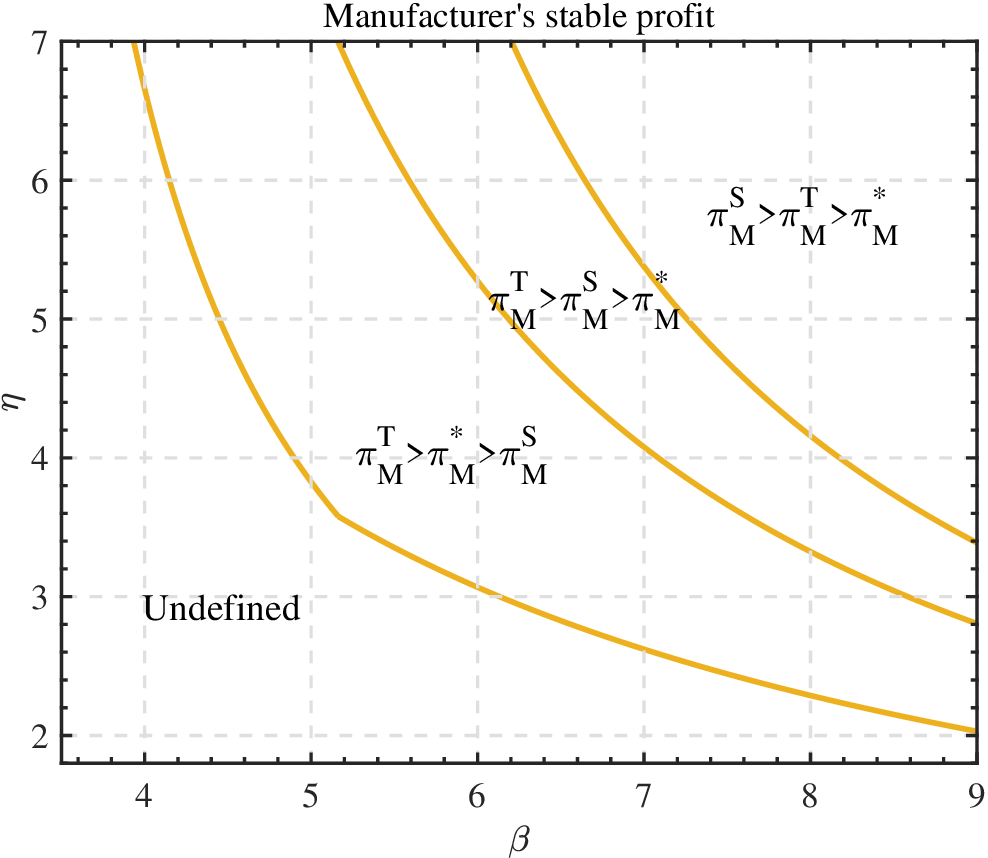}
    \caption{Comparisons between $\pi_G(\infty)$ and $\pi_M(\infty)$.}
    \label{fig:4}
    \end{minipage}
    \begin{minipage}{0.325\linewidth}
    \centering
    \vspace{1.8mm}
    \includegraphics[scale = 0.3]{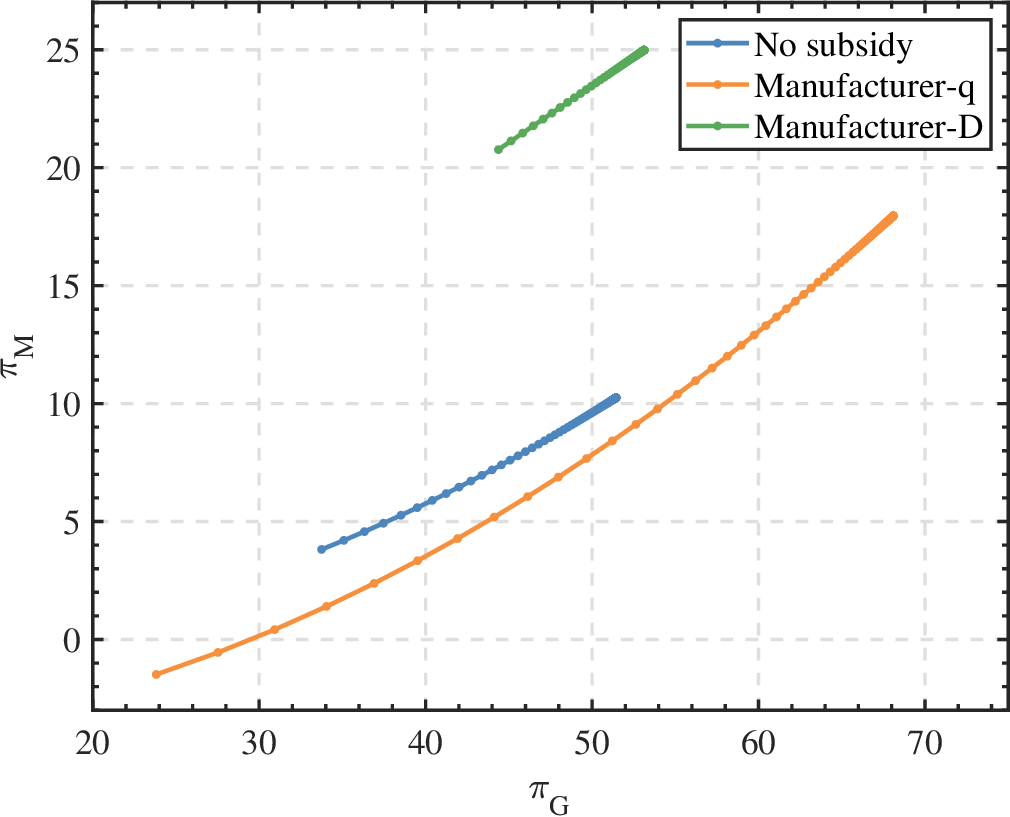}
    \caption{The dynamics of $\pi_G$ and $\pi_M$.}
    \label{fig:5}
    \end{minipage}
    \put(-430,50){\normalsize{(a)}}
    \put(-285,50){\normalsize{(b)}}
    \put(-80,27){\LARGE{$\nearrow$}}
\end{figure}

Governments typically adopt subsidy policies aimed at maximizing their revenue. Proposition \ref{prop:9} outlines the criteria that a government should consider when selecting an appropriate subsidy strategy under specific conditions. To illustrate this concept in a general context, a numerical example is presented in Figure \ref{fig:4}(a). The figure clearly demonstrates that when customer price sensitivity or the government's marginal sales revenue is relatively high, as indicated by the upper-right corner of the figure, Policy (S) is more advantageous than Policy (T). In contrast, in other scenarios, Policy (T) is the preferred option.

To further explore the stable profit for the manufacturer, a numerical example is presented in Figure \ref{fig:4}(b). A visual comparison of the figures reveals similar shapes, indicating that, for most values of $\beta$ and $\eta$, the optimal subsidy policy for the manufacturer aligns closely with that of the government. This alignment suggests potential implicit cooperation between the government and the manufacturer.

Figure \ref{fig:5} illustrates the dynamic relationship between government profit and manufacturer profit, with each point representing a unit of time. The graph shows a positive correlation between government and manufacturer profits for both subsidy policies, reinforcing the implicit cooperative bond between the two entities. Specifically, as government profit increases, so does the manufacturer's profit. Notably, under Policy (T), despite initially low profits for both parties—especially due to high technology investments—the manufacturer's profit may even be negative. However, the growth rate of profits is rapid. When comparing the three scenarios, Policy (T) results in the highest overall stable profits for both the government and the manufacturer.

\begin{figure}[!h]
    \centering
    \includegraphics[scale = 0.3]{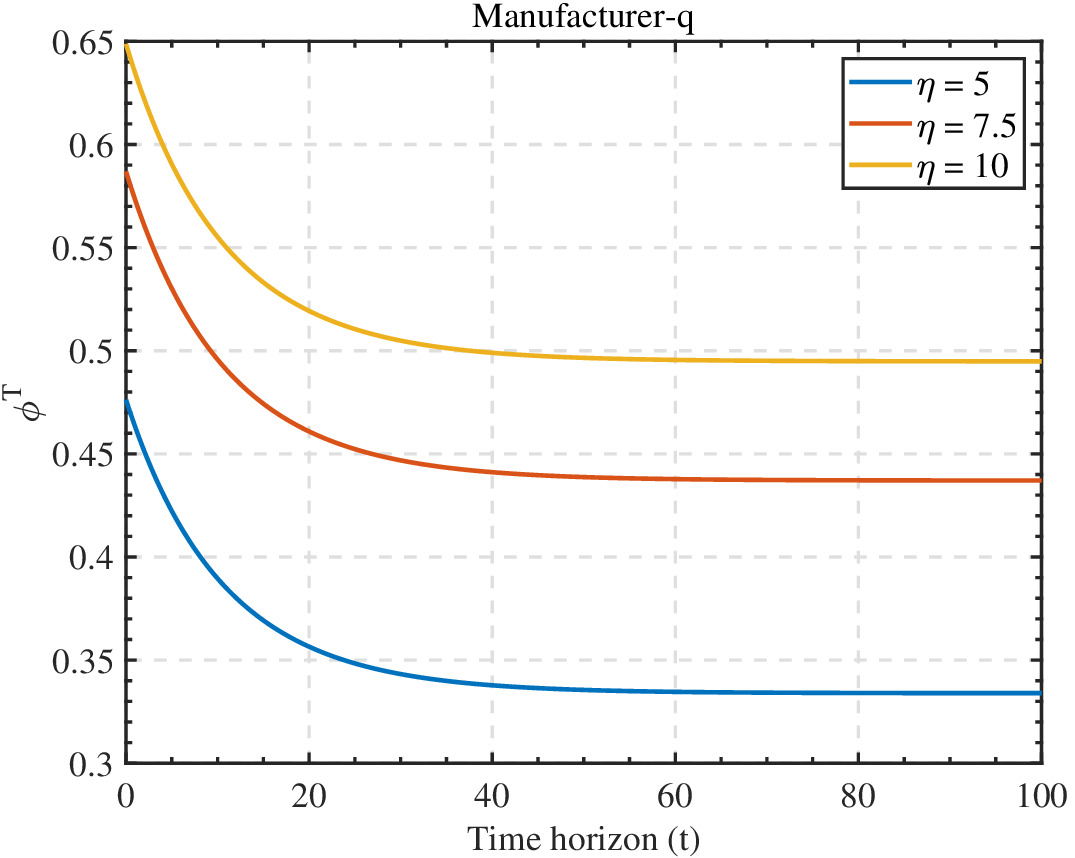}
    \qquad
    \includegraphics[scale = 0.3]{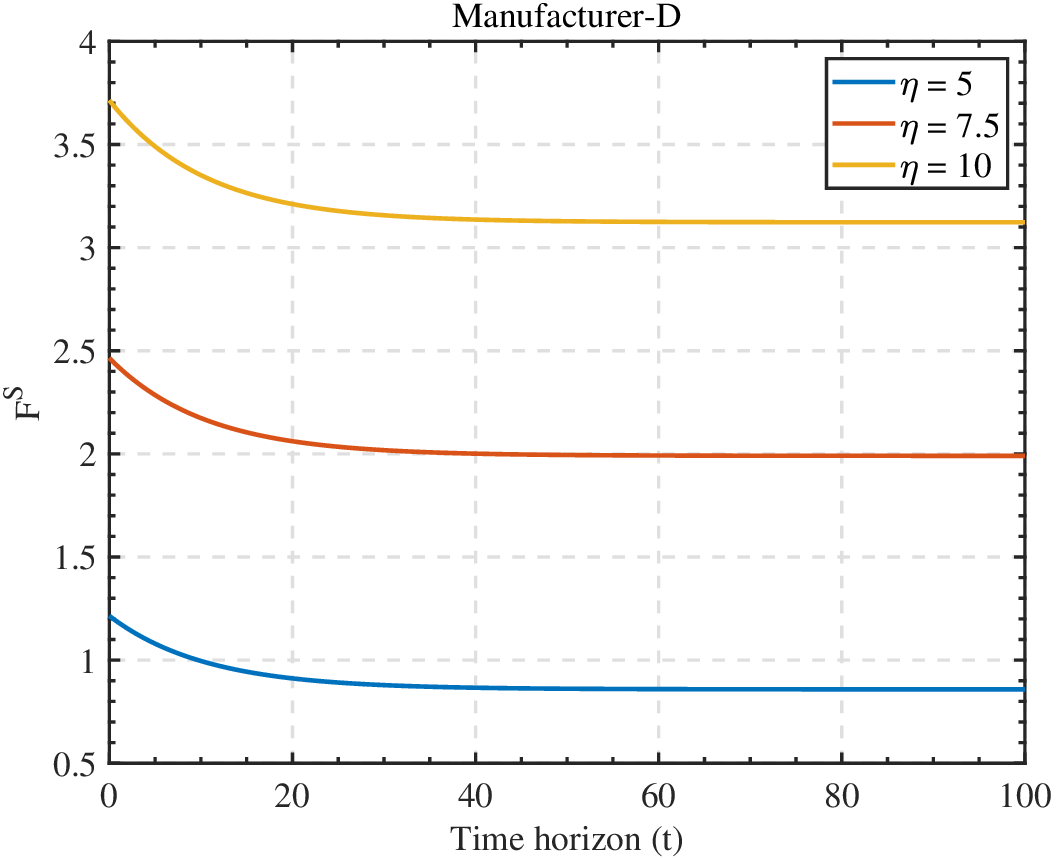}
    \caption{Equilibrium controls of $\phi^T$ and $F^S$ under two types of subsidy policies.}
    \label{fig:6}
\end{figure}

\begin{figure}[!h]
    \centering
    \includegraphics[scale = 0.3]{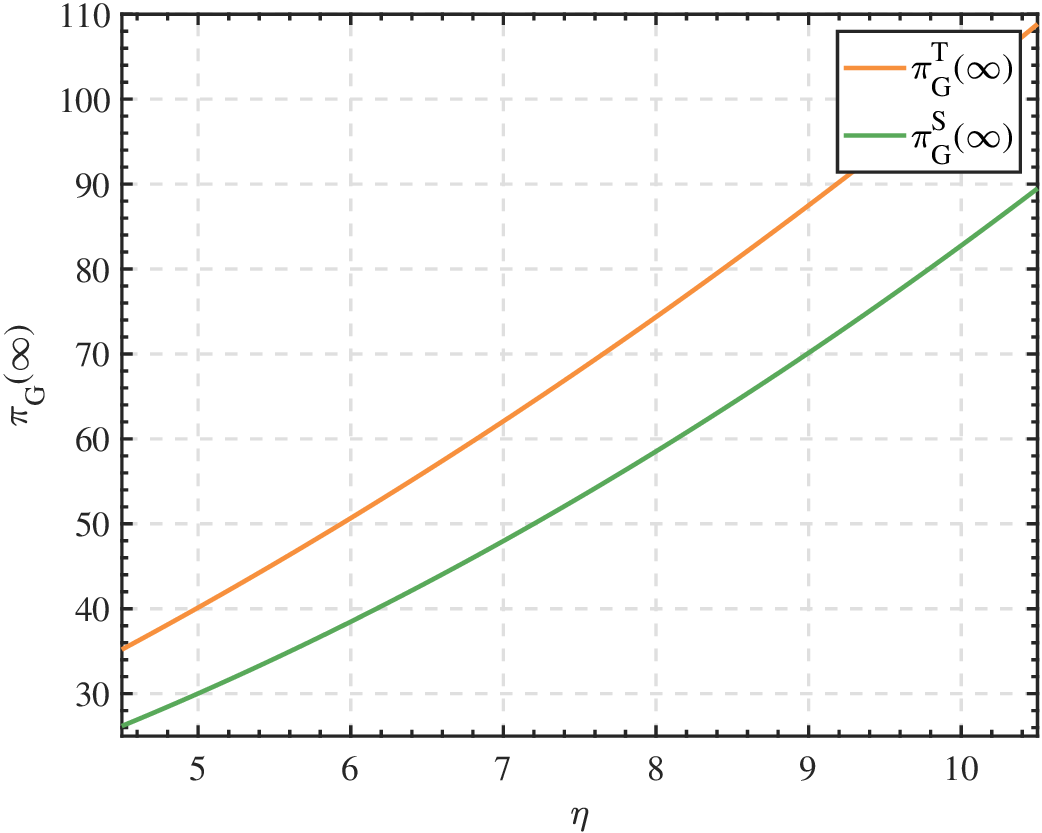}
    \qquad
    \includegraphics[scale = 0.3]{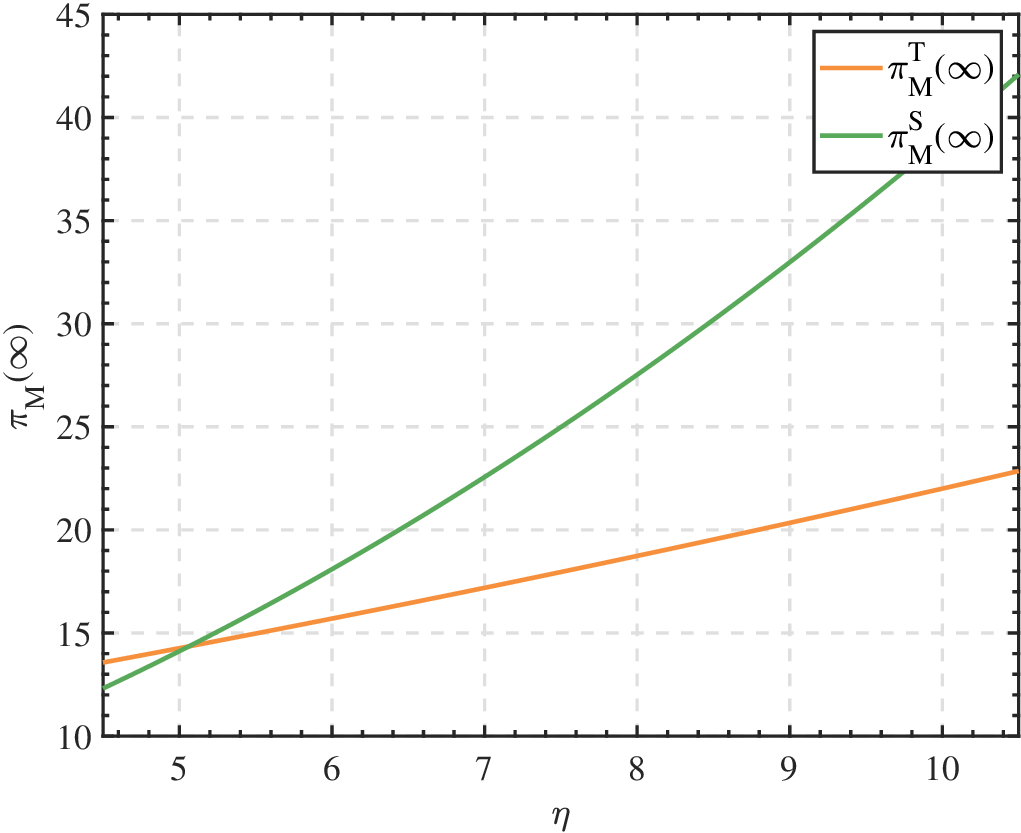}
    \caption{The impact of marginal sales revenue on $\pi_G(\infty)$ and $\pi_M(\infty)$.}
    \label{fig:7}
\end{figure}

Considering the effect of the government's marginal revenue on the cooperative relationship with the manufacturer, Figure \ref{fig:6} shows that, regardless of the chosen subsidy policy, the government's subsidy to the manufacturer decreases over time, consistent with Proposition \ref{prop:4}. Additionally, as the government's marginal revenue increases, its subsidy to the manufacturer rises. In Figure \ref{fig:7}, an increase in the government's marginal revenue leads not only to an increase in government profit but also to an increase in manufacturer profit. Notably, the growth rate of the manufacturer's profit under Policy (S) is significantly higher than that under Policy (T). This pattern suggests that the government is cooperating with the manufacturer by sharing marginal revenue, thereby enhancing the efficiency of the vaccine supply chain.

Based on these findings, several key recommendations can be made. First, governments should carefully evaluate customer price sensitivity and their own marginal sales revenue when selecting an optimal subsidy policy. If both consumer price sensitivity and government marginal revenue are high, Policy (S) should be prioritized to generate higher short-term sales and immediate profits. However, for long-term sustainability and enhanced product quality, Policy (T) may offer more substantial benefits. Manufacturers should align their strategies with the government's chosen policy to maximize profits through implicit cooperation, particularly in contexts where long-term investments are crucial. In cases where the government adopts Policy (T), manufacturers should prepare for initially lower profits due to higher technology investments but should capitalize on future profit potential as sales volumes increase.

\begin{figure}[!h]
    \centering
    \includegraphics[scale = 0.245]{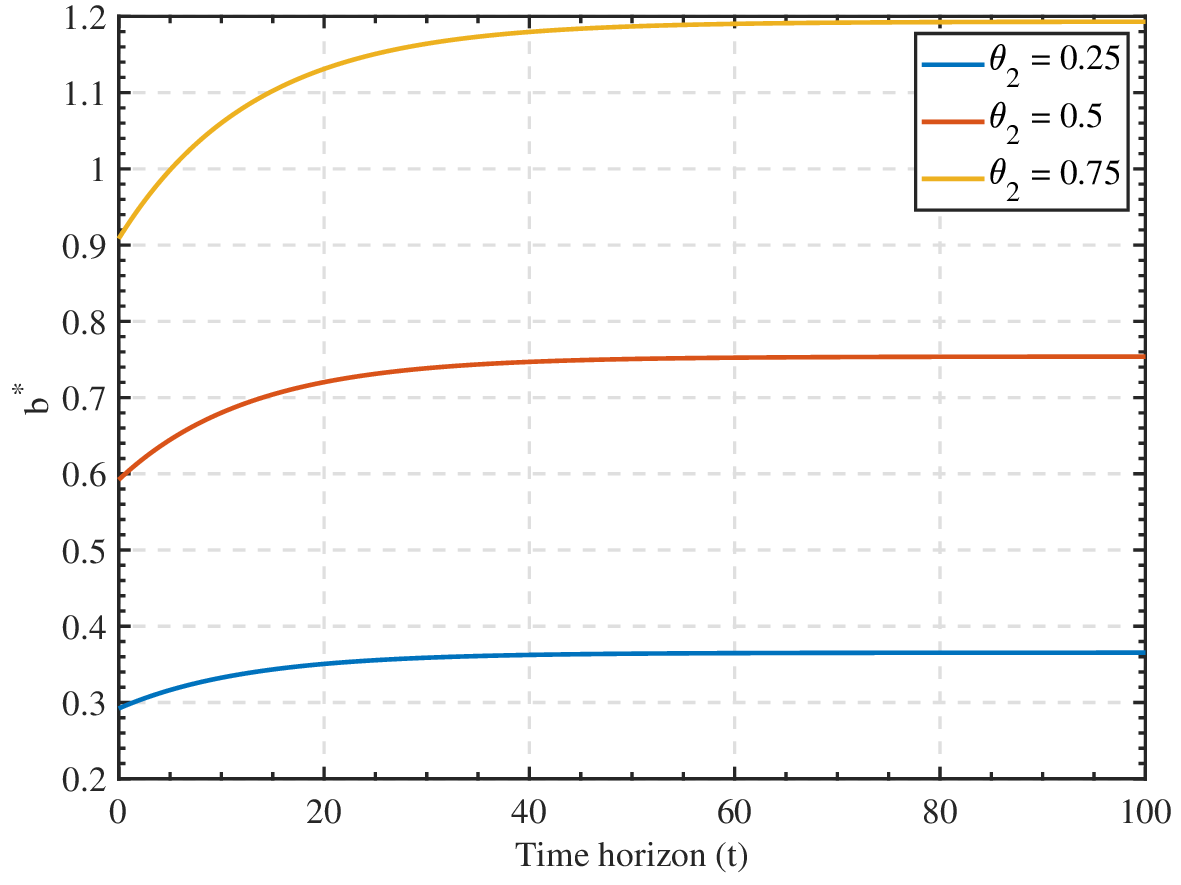}
    \quad
    \includegraphics[scale = 0.245]{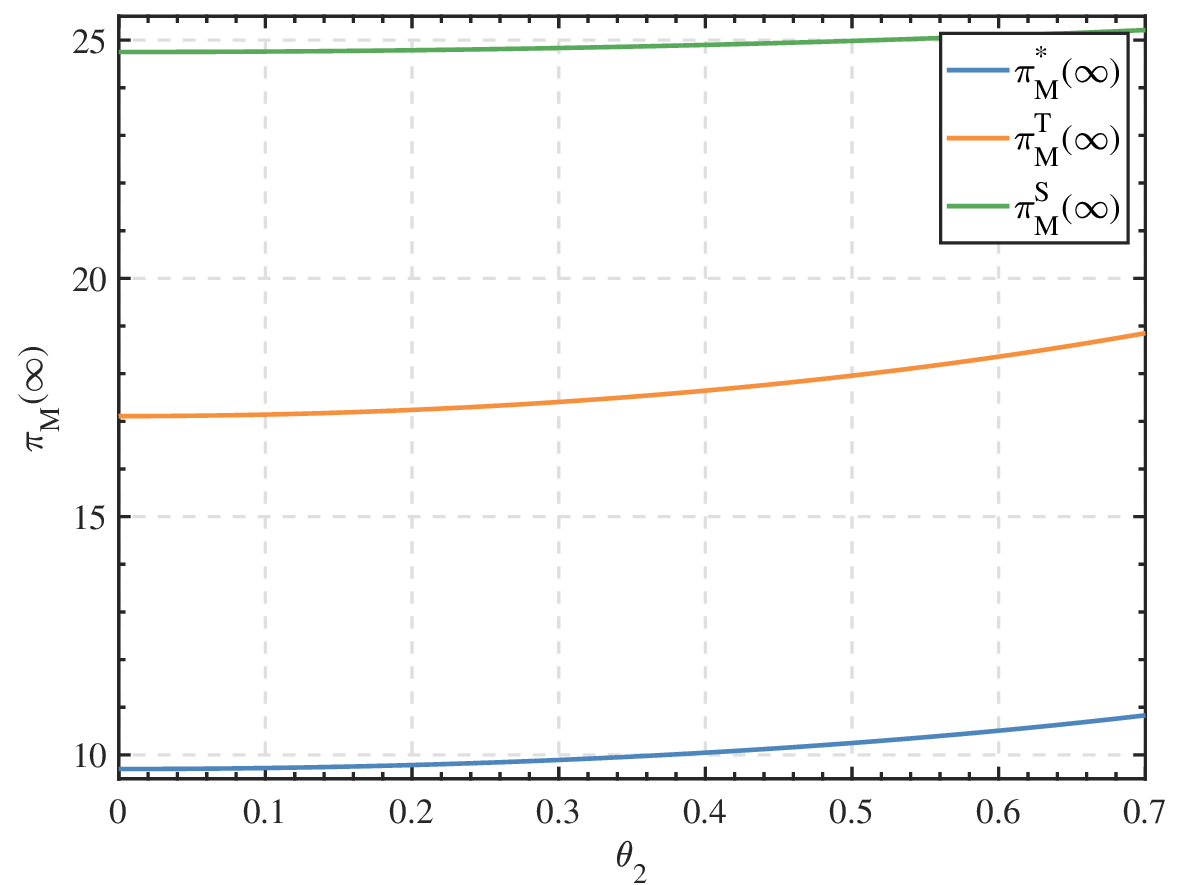}
    \quad
    \includegraphics[scale = 0.245]{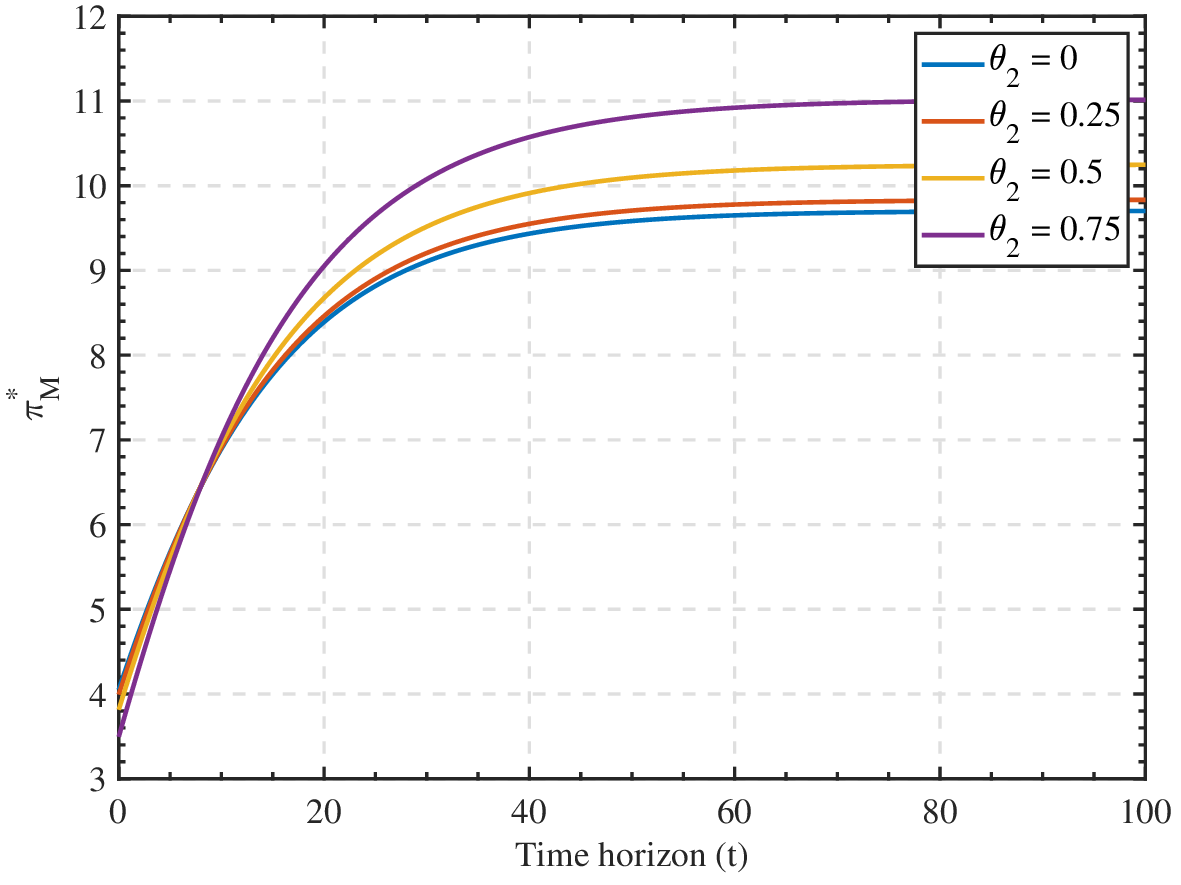}
    \caption{The impact of blockchain technology on $b$ and $\pi_M$.}
    \label{fig:8}
\end{figure}

\subsection{Impact of Blockchain and Key Parameters}

In general, the greater the manufacturer's investment in blockchain technology, the more robust the blockchain system becomes, which, in turn, enhances the level of goodwill. As illustrated in Figure \ref{fig:8}, in the absence of subsidies, the manufacturer's investment in blockchain technology increases over time. Furthermore, as $\theta_2$ rises, the manufacturer's involvement in blockchain technology becomes more pronounced. 

Regarding the manufacturer's profits, blockchain technology contributes to stabilizing long-term profits, with the growth rate accelerating as $\theta_2$ increases. Policy (T) further incentivizes the manufacturer to invest in blockchain technology, leading to the highest growth rate in profits. Ultimately, while the manufacturer's initial profits may decline significantly at high levels of $\theta_2$, profit growth in the later stages becomes more substantial. In practice, the adoption of blockchain technology can enhance the manufacturer's cumulative profits.

\begin{figure}[!h]
    \centering
    \includegraphics[scale = 0.28]{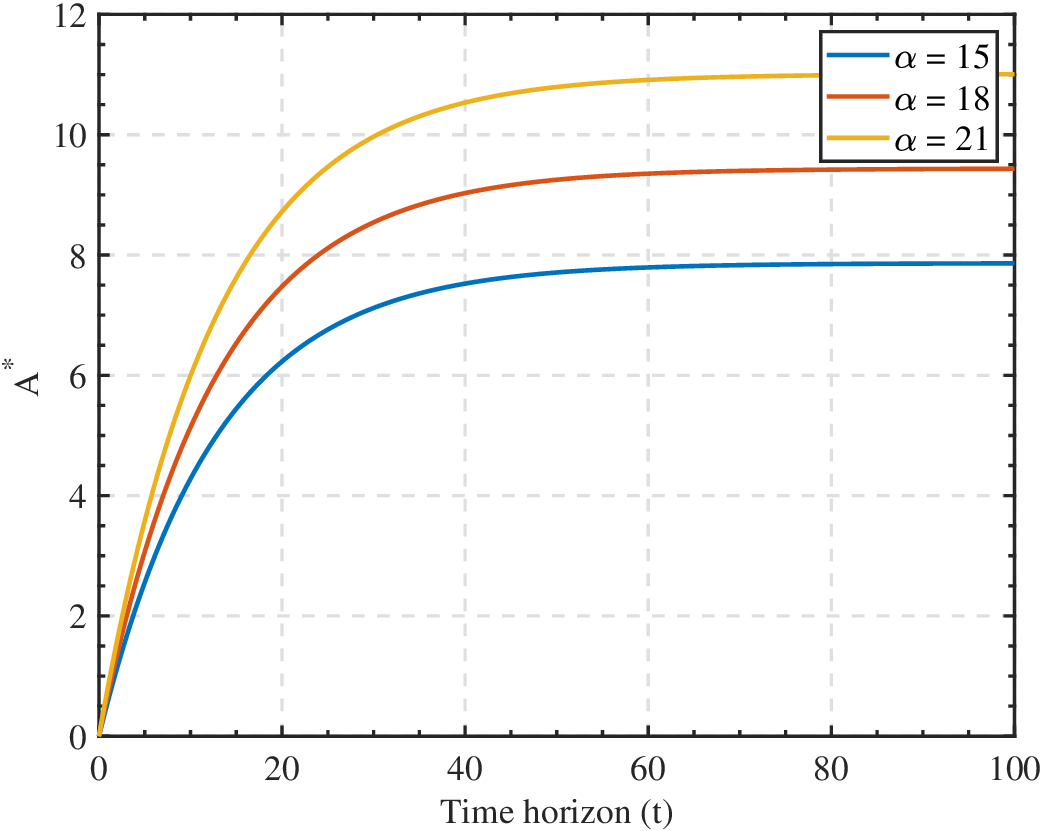}
    \quad
    \includegraphics[scale = 0.28]{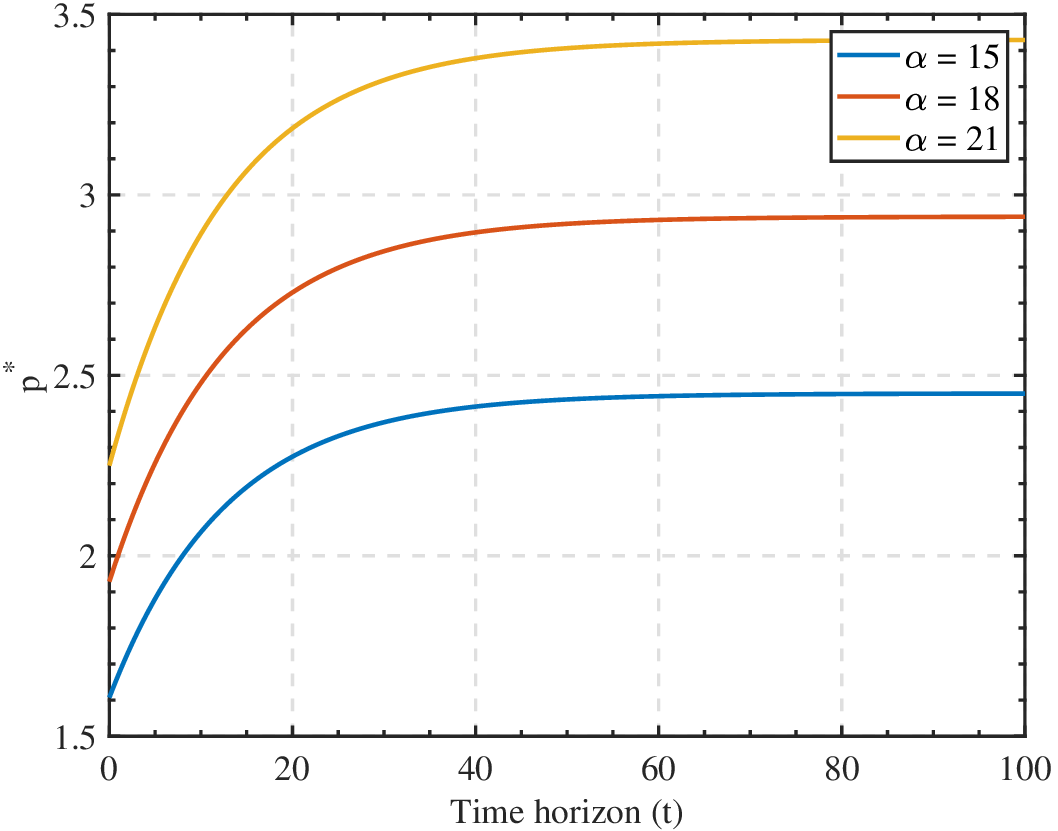}
    \quad
    \includegraphics[scale = 0.28]{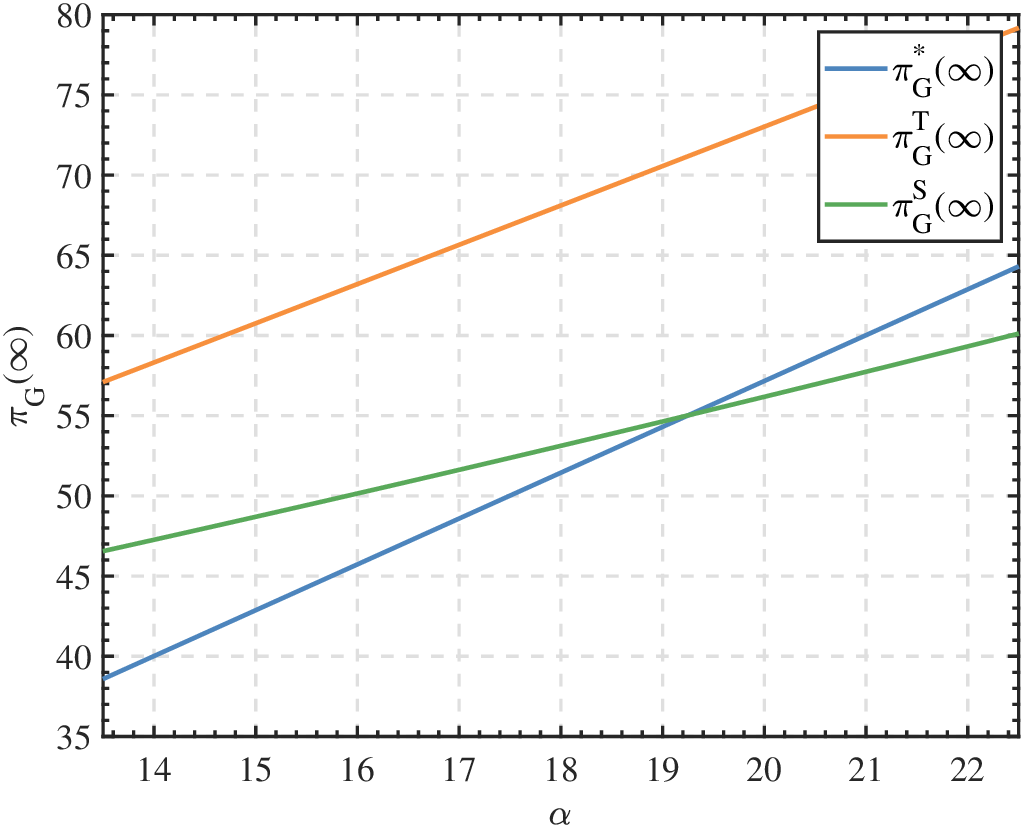}
    \caption{The impact of $\alpha$ on $A$, $p$, and $\pi_G(\infty)$.}
    \label{fig:9}
\end{figure}

Now, consider the impact of market capacity. As shown in Figure \ref{fig:9}, an increase in market capacity improves both the quality and goodwill levels of vaccines, resulting in higher vaccine prices in the absence of subsidies. Moreover, regardless of the subsidy policy implemented by the government, the government's steady profits increase. However, when the government adopts Policy (S), the growth rate of government profits is particularly slow. This is because, under Policy (S), the manufacturer and retailer stimulate sales by lowering prices. For larger market volumes, such price reductions force the manufacturer and retailer to absorb more hidden costs, which negatively affects vaccine sales.

\begin{figure}[!h]
    \centering
    \includegraphics[scale = 0.27]{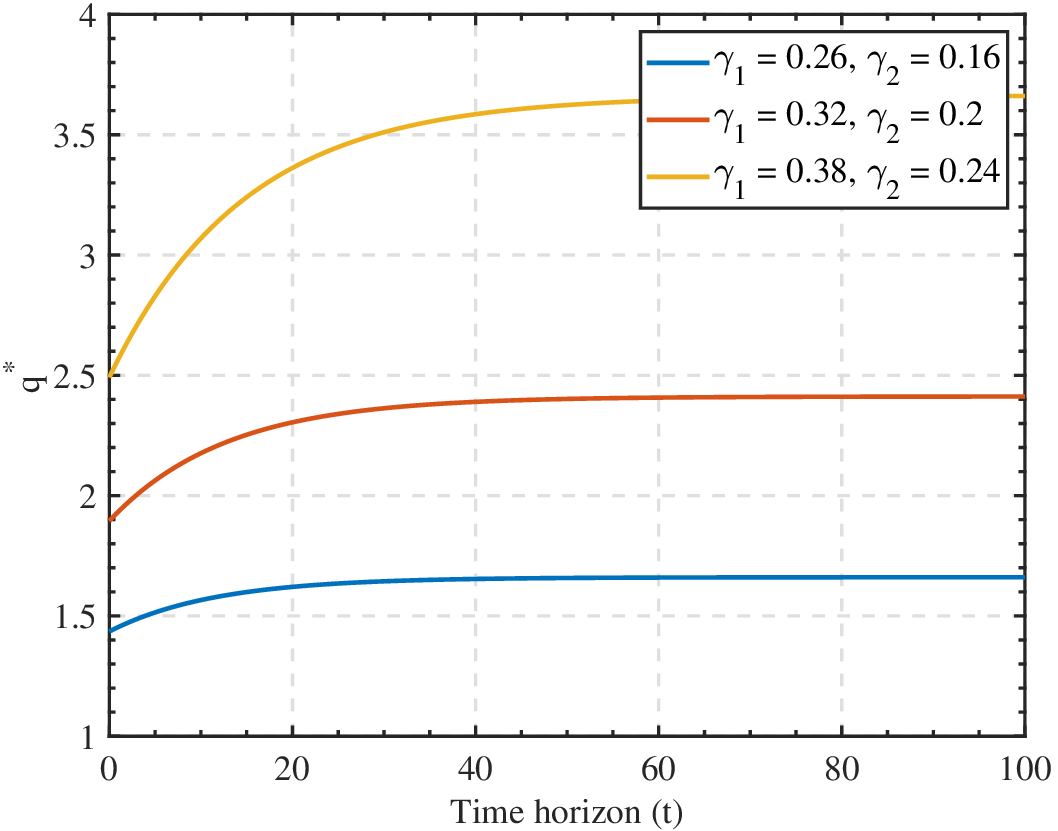}
    \quad
    \includegraphics[scale = 0.27]{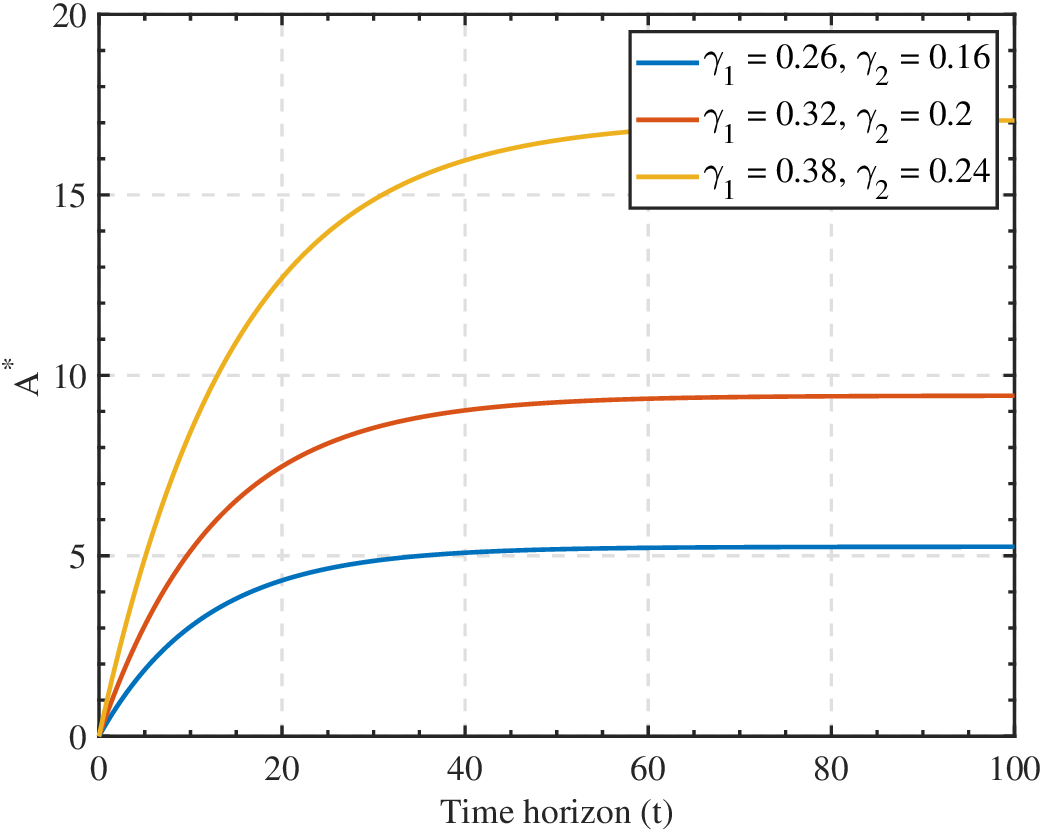}
    \quad
    \includegraphics[scale = 0.255375]{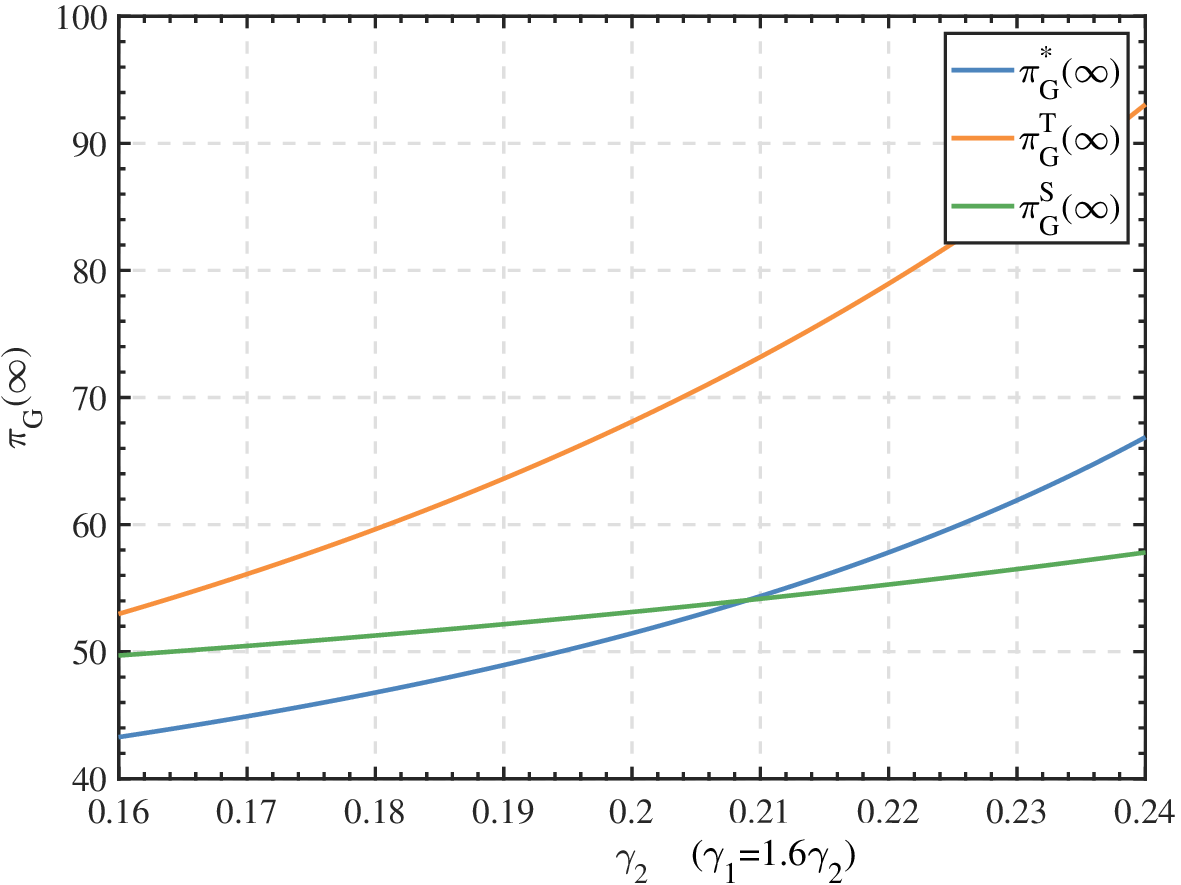}
    \caption{The impacts of $\gamma_1$ and $\gamma_2$ on $q$, $A$, and $\pi_G(\infty)$.}
    \label{fig:10}
\end{figure}

Finally, considering the effects of quality and goodwill levels, Figure \ref{fig:10} illustrates that, in the absence of subsidies, as both $\gamma_1$ and $\gamma_2$ increase, the manufacturer raises its investment in technology to improve the quality and goodwill of the vaccines, thereby meeting market demand. This behavior mirrors the manufacturer's blockchain investment and the retailer's advertising expenditure. Moreover, when $\gamma_1$ and $\gamma_2$ rise proportionally, the government's stable profit increases, regardless of the chosen subsidy policy. However, Policy (T) accelerates the growth rate of the government's profit, as it subsidizes the manufacturer's technology investment, aligning with prevailing market trends. In contrast, Policy (S) results in slower growth due to the reduced investment intensity by both the manufacturer and retailer, which contrasts with market trends.

\section{Managerial Insights}
\label{sec:mi}

This paper investigates the characteristics and dynamic behavior of a three-tier vaccine supply chain comprising the government, manufacturer, and retailer, with government support. It offers valuable insights to guide decision-making regarding sales, investments, and collaboration among stakeholders in the vaccine supply chain. Specifically, the paper addresses the following key issues:

(1) Evaluation of Government Subsidy Policies: The paper analyzes three different government subsidy policies. First, it considers the subsidy policy of proportional reimbursement for vaccination costs, highlighting the necessity for effective cooperation between the government, manufacturer, and retailer to ensure success. It then compares the relationships among key variables under Policy (T), Policy (S), and the no-subsidy scenario. Policy (T) focuses on boosting vaccine quality and goodwill through significant, sustained investments in technology and advertising by the manufacturer and retailer, leading to higher future profits for the government, thus representing a forward-looking strategy. In contrast, Policy (S) seeks to stimulate immediate vaccine sales through continuous price reductions, resulting in higher short-term profits for the government, making it a more short-sighted strategy. From a managerial perspective, we recommend that governments prioritize long-term outcomes and adopt Policy (T), whereas Policy (S) may be more appropriate for scenarios prioritizing immediate benefits, such as during a pandemic outbreak. Additionally, the paper offers guidance on selecting the optimal subsidy policy based on different contexts. For instance, if the public is highly price-sensitive and the per capita vaccination benefit is substantial, Policy (S) should be favored; otherwise, Policy (T) is preferable. This distinction is crucial for policymakers seeking to balance short-term public health needs with long-term sustainability.

(2) Analysis of Optimal Decision Strategies under Various Subsidy Policies: The paper proposes optimal subsidy strategies for the government, along with pricing and investment strategies for the manufacturer and retailer within each government subsidy framework. It provides best practices for decision-making while also analyzing the characteristics of optimal strategies from multiple perspectives. Over time, with ample profits and market expansion, manufacturers' and retailers' investments should increase, while government subsidies should gradually decrease as the vaccine supply chain matures. In static terms, for example, in scenarios where the market size is large and the public places more value on vaccine quality and reputation than on price sensitivity, manufacturers should focus on increasing technological investments. These insights contribute to a deeper understanding of optimal decision-making logic, enabling timely adjustments in response to market changes and improving the overall efficiency and adaptability of the vaccine supply chain.

(3) Potential Impact of Blockchain Technology on the Vaccine Supply Chain: This paper explores the potential benefits of integrating blockchain technology into the vaccine supply chain, including improvements in vaccine quality and goodwill, as well as increased future profits for manufacturers. We recommend that manufacturers incorporate blockchain technology to strengthen the supply chain. Through numerical experiments, the paper demonstrates how blockchain can enhance the supply chain and offers guidance to manufacturers and retailers on adjusting strategies in response to market changes, such as increased investment and higher prices. The integration of blockchain can also improve transparency and traceability, which are crucial for maintaining public trust during vaccination campaigns.

(4) Exploration of Government-Manufacturer Collaboration: This paper examines the implicit cooperation between the government and manufacturers in the vaccine supply chain, exploring opportunities for mutual benefit. It focuses on the alignment of optimal subsidy policies for both parties, particularly when the government's marginal benefit is substantial. In such cases, the profits of both the government and manufacturer grow in tandem, and as the government's marginal benefit increases, its subsidy to the manufacturer also rises. We encourage various forms of cooperation between the government and manufacturers to strengthen their relationship, ensuring the stability and sustainability of the vaccine supply chain. This collaborative approach can help mitigate risks associated with supply chain disruptions and improve overall effectiveness.

Given these insights, several practical recommendations emerge for improving the management of the vaccine supply chain. Governments should prioritize long-term outcomes when selecting subsidy policies, especially when aiming to achieve sustainable improvements in vaccine quality and public goodwill. This requires investments in long-term strategies such as Policy (T), which supports future profitability and market stability. In contrast, during urgent public health crises, such as a pandemic, Policy (S) provides an effective short-term solution. Manufacturers and retailers should align their strategies with government policies to maximize both long-term profitability and supply chain efficiency. In particular, manufacturers should consider increasing their technological investments when market conditions prioritize quality over price sensitivity. Blockchain technology presents a promising opportunity to enhance trust and improve the overall supply chain, making it a critical tool for manufacturers. Lastly, fostering a collaborative relationship between the government and manufacturers is essential. By sharing marginal benefits and aligning their objectives, both parties can create a mutually beneficial environment that strengthens supply chain resilience and improves vaccine distribution efficiency. This comprehensive approach ensures that the vaccine supply chain remains adaptable to both current and future public health needs.

\section{Conclusion and Discussion}
\label{sec:c}

This paper examines a three-tier vaccine supply chain consisting of the government, the manufacturer, and the retailer. We employ dynamic differential equations to model the quality and goodwill levels of vaccines, which leads to the development of a differential game model. This model incorporates pricing, advertising investment, blockchain technology, and three distinct government subsidy policies, including proportional subsidies for technology investments and volume-based subsidies. Within the differential game framework, the government, as the leader, initially determines its subsidy intensity, influencing the manufacturer's decisions regarding technology investment, blockchain adoption, and wholesale prices. Subsequently, the retailer makes decisions about advertising investments and retail prices. Under the open-loop assumption, we apply Pontryagin's Maximum Principle to derive the optimal strategies for the government, manufacturer, and retailer under each subsidy policy across different scenarios. Through analysis and numerical simulations, we arrive at several key findings, which are summarized as follows:

\begin{enumerate}
\item In the absence of cooperation among the government, vaccine manufacturers, and retailers, a subsidy policy that proportionally reimburses vaccination costs to vaccinated individuals is ineffective.
\item For the government, Policy (T) represents a forward-looking strategy, whereas Policy (S) is more short-sighted.
\item When the government's marginal sales revenue $\eta$ or price sensitivity $\beta$ is relatively high, the government tends to favor Policy (S); otherwise, it tends toward Policy (T).
\item Blockchain technology positively impacts the vaccine supply chain, particularly by enhancing the late-stage profits of manufacturers.
\item An implicit mutual benefit exists between the government and the manufacturer, fostering a win-win cooperation dynamic.

\end{enumerate}

The study highlights several avenues for future research. Future work could refine the model assumptions to enhance its adaptability. Moreover, while the current model simplifies sales as a linear combination of retail price, quality, and goodwill, it overlooks other critical factors, such as vaccine accessibility. Addressing these factors will require more advanced modeling techniques. To increase practical relevance, future studies could incorporate multiple manufacturers and retailers into the vaccine supply chain model, enabling the exploration of both competitive and cooperative dynamics. Additionally, incorporating stochastic elements, such as random errors in quality and goodwill, and employing robust optimization methods could enhance model accuracy. Furthermore, variables such as cold chain logistics and production losses should be considered to provide a more comprehensive view of supply chain operations. Finally, empirical studies validating and evaluating the theoretical findings would be a valuable direction for future research.

\theendnotes

\section*{Disclosure Statement}
The authors report there are no competing interests to declare.

\section*{Data Availability Statement}
Data sharing is not applicable to this article as no new data were created or analyzed in the study.

\appendix
\section{Supplemental Online Material}
\label{app}

Supplemental online material for this article can be found online at (TBD).

\bibliographystyle{apalike} 
\bibliography{vaccine}

\end{document}